\documentclass[sigconf]{aamas} 

\usepackage{balance} 

\definecolor{mydarkblue}{rgb}{0,0.08,0.45}
\usepackage{hyperref}
\hypersetup{ %
    colorlinks=true,
    linkcolor=mydarkblue,
    citecolor=mydarkblue,
    filecolor=mydarkblue,
    urlcolor=mydarkblue,
    pdfview=FitH}
\usepackage{url}
\usepackage[utf8]{inputenc} 
\usepackage[T1]{fontenc}    

\usepackage{mysymbols}
\usepackage{booktabs}       
\usepackage{algorithm}
\usepackage{algpseudocode}

\usepackage{amsmath}
\usepackage{amsthm}
\usepackage{attrib}
\usepackage{stmaryrd}
\usepackage{latexsym}
\usepackage{bm}
\usepackage{xcolor}
\usepackage{subcaption}
\usepackage{placeins}
\theoremstyle{plain}
\usepackage{mathtools}
\newcommand{\defeq}{\vcentcolon=}
\usepackage{thmtools,thm-restate}
\usepackage{cleveref}
\usepackage{wrapfig}
\usepackage{enumitem}
\usepackage[normalem]{ulem}
\newcommand{\stkout}[1]{\ifmmode\textcolor{red}{\text{\sout{\ensuremath{#1}}}}\else\sout{#1}\fi}
\DeclareMathOperator*{\argmax}{argmax}
\DeclareMathOperator*{\argmin}{argmin}
\usepackage{tikz}
\usetikzlibrary{automata, positioning, arrows}

\tikzset{
    ->, 
    >=stealth,
}

\renewenvironment{quote}{%
  \list{}{%
    \leftmargin1.0cm   
    \rightmargin\leftmargin
  }
  \item\relax
}
{\endlist}

\definecolor{bettergreen}{HTML}{4faa42}
\definecolor{mygreen}{HTML}{91ed92}
\definecolor{myred}{HTML}{f28d8d}
\definecolor{mygrey}{HTML}{a1a1a1}

\setcopyright{ifaamas}
\acmConference[AAMAS '22]{Proc.\@ of the 21st International Conference
on Autonomous Agents and Multiagent Systems (AAMAS 2022)}{May 9--13, 2022}
{Online}{P.~Faliszewski, V.~Mascardi, C.~Pelachaud,
M.E.~Taylor (eds.)}
\copyrightyear{2022}
\acmYear{2022}
\acmDOI{}
\acmPrice{}
\acmISBN{}

\acmSubmissionID{199}

\title[Adaptive Incentive Design with Multi-Agent Meta-Gradient Reinforcement Learning]{Adaptive Incentive Design with Multi-Agent Meta-Gradient Reinforcement Learning}

\author{Jiachen Yang\textsuperscript{\normalfont 1},
Ethan Wang\textsuperscript{\normalfont 1}, 
Rakshit Trivedi\textsuperscript{\normalfont 2}, 
Tuo Zhao\textsuperscript{\normalfont 1} \& 
Hongyuan Zha\textsuperscript{\normalfont 3}}
\affiliation{
\institution{
\textsuperscript{1}Georgia Institute of Technology,
\textsuperscript{2}Harvard University,
\textsuperscript{3}Chinese University of Hong Kong, Shenzhen}
\city{}\country{}}
\email{
jyang462,ewang78,tourzhao@gatech.edu,rstrivedi@seas.harvard.edu,zhahy@cuhk.edu.cn}

\begin{abstract}
Critical sectors of human society are progressing toward the adoption of powerful artificial intelligence (AI) agents, which are trained individually on behalf of self-interested principals but deployed in a shared environment. Short of direct centralized regulation of AI, which is as difficult an issue as regulation of human actions, one must design institutional mechanisms that indirectly guide agents' behaviors to safeguard and improve social welfare in the shared environment. Our paper focuses on one important class of such mechanisms: the problem of adaptive incentive design, whereby a central planner intervenes on the payoffs of an agent population via incentives in order to optimize a system objective. To tackle this problem in high-dimensional environments whose dynamics may be unknown or too complex to model, we propose a model-free meta-gradient method to learn an adaptive incentive function in the context of multi-agent reinforcement learning. Via the principle of online cross-validation, the incentive designer explicitly accounts for its impact on agents' learning and, through them, the impact on future social welfare. Experiments on didactic benchmark problems show that the proposed method can induce selfish agents to learn near-optimal cooperative behavior and significantly outperform learning-oblivious baselines. When applied to a complex simulated economy, the proposed method finds tax policies that achieve better trade-off between economic productivity and equality than baselines, a result that we interpret via a detailed behavioral analysis.
\end{abstract}


\newcommand{\BibTeX}{\rm B\kern-.05em{\sc i\kern-.025em b}\kern-.08em\TeX}

\begin{document}


\pagestyle{fancy}
\fancyhead{}


\maketitle 


\begin{quote}
  Madness is rare in individuals---but in groups, parties, nations, and ages, it is the rule.
  \begin{flushright} \citet{nietzsche1989beyond}
  \end{flushright}
\end{quote}

\section{Introduction}
\label{sec:introduction}

As advances in artificial intelligence research drive the growing presence of AI in critical infrastructures---such as transportation \citep{bonnefon2016social,bissell2020autonomous}, information communication \citep{howard2018algorithms}, financial markets \citep{hsu2016bridging} and agriculture \citep{liakos2018machine}---it is increasingly important for AI research to complement the solipsistic view of  models and agents acting in isolation with a broader viewpoint: these models and agents may be developed by independent and self-interested principals but are eventually deployed in a shared multi-agent ecosystem.
Apart from the special cases of pure common or conflicting interest (i.e., team or zero-sum payoffs), the majority of possible scenarios involve mixed motives \citep{robinson2005topology} whereby cooperation for optimal group payoff is attainable if selfish behavior out of greed or fear---which results in low individual and group payoff---can be overcome.
Research on endowing individual agents with social capabilities and designing central institutional mechanisms to safeguard and improve social welfare, recently named ``Cooperative AI'' \citep{dafoe2020open}, is a long-term necessity that requires present-day research efforts.

We focus on one specific pillar of this research agenda: the problem of \textit{adaptive incentive design} \citep{ratliff2019perspective}, whereby a central institution shapes the behavior of self-interested agents to improve social welfare, by introducing an incentive function to modify their individual payoffs.
The trivial solution of setting individual payoffs equal to the average system payoff is known to be suboptimal \citep{paccagnan2019utility}: e.g., uniform redistribution of income leads to low productivity.
This problem can be formalized as a reverse Stackelberg game \citep{ho1981information,stackelberg1934}, in which the leader first proposes a function (e.g., the incentive function) that maps from the follower's action space to the leader's decision space, 
while the follower chooses a best response.
This is a difficult bi-level optimization problem even in the linear case \citep{groot2012reverse}.
To tackle this problem in high-dimensional multi-agent systems, we propose a model-free method based on meta-gradient reinforcement learning \citep{xu2018meta} and the principle of online cross-validation \citep{sutton1992adapting}, 
for the incentive designer (ID) to explicitly account  for the learning dynamics of agents in response to incentives.
Potential applications in the long term include: e.g. shaping consumption patterns to improve the efficiency of smart power grids \citep{vardakas2014survey} and to mitigate climate crisis \citep{barfuss2020caring}; reducing wait times or traffic congestion in taxi dispatch and traffic tolling systems \citep{de2011traffic,miao2016taxi}; solving the social dilemma of autonomous vehicle adoption \citep{bonnefon2016social}; improving the trade-off between economic productivity and income equality via taxation \citep{zheng2020ai}.


In the spirit of complexity economics \citep{arthur2021foundations}, whereby the tractability of linear models with analytic equilibria \citep{gopalakrishnan2014potential,ratliff2020adaptive,paccagnan2019utility} is eschewed for the greater realism and richer dynamics of high-dimensional agent-based simulation, we work in the framework of Markov games \citep{littman1994markov} with reinforcement learning (RL) agents \citep{sutton2018reinforcement}.
Within this context, we interpret ``incentive design'' in the broad sense of influencing agents' behaviors via modifying their individual payoffs.
The issue of incentive compatibility, despite being central to analytically tractable applications such as auctions where the goal is to elicit truthful reporting of private valuations \citep{ratliff2019perspective}, do not pertain in general to complex simulations involving RL agents for the following reasons: 1) the concept of private individual preferences may not make sense in the application (e.g., a social dilemma whose payoff is known to all parties); 2) there may be no \textit{a priori} or \textit{fixed} private valuations, because an agent's preference is completely represented by its reward function, which itself depends on the incentive function and hence changes dynamically along with the ID's online optimization process
; and 3) a complex simulation involving nonlinear processes and discrete rules is used to investigate dynamical behavior rather than to converge to equilibria.

Our use of simulation is motivated by two considerations.
Firstly, there may arise a future ecosystem of AI 
in diverse spaces, such as multiple firms in a financial market \citep{ammanath2020thriving} and multiple recommendation systems in the same consumer sector \citep{lu2015recommender}.
With the increasing success of RL on progressively more difficult tasks \citep{silver2017mastering,vinyals2019grandmaster,berner2019dota} and growing efforts to apply RL to real-world problems \citep{lazic2018data,ie2019recsim}, such real-world \textit{in silico} AI are likely to involve RL for optimization of long-term objectives and will be ontologically equivalent to the entities in our work.
Secondly, agent-based simulation is also relevant to incentive design for a group of self-interested humans, firms, or states, by viewing the reward-maximizing behavior of RL agents as an approximation of the bounded rationality of such real world entities \citep{niv2009reinforcement}.
Therefore, we validate our approach in existing simulation benchmarks: \textit{Escape Room} \citep{yang2020learning} and \textit{Cleanup} \citep{hughes2018inequity}, which are deceptively hard but easily interpretable benchmark problems, and the complex economic simulation called \textit{Gather-Trade-Build} \citep{zheng2020ai}, where agent behavior such as specialization and correlations between taxation and productivity are qualitatively in accord with results from economic theory and reality.

The centralized incentive designer (ID) in our work is related but orthogonal to the centralization commonly seen in the literature on cooperative MARL \citep{panait2005cooperative,oroojlooyjadid2019review}: centralized training in MARL permits a global entity to update each agent's individual policy parameters using shared global information and thereby directly optimize a single team reward, while the ID in our work can only set an incentive function to affect each agent's total individual reward.
Given a fixed incentive function, our agents live in a fully-decentralized POMDP \citep{oliehoek2016concise}.
The ID's intervention on the agents' reward function also differs from the insertion of controlled agents (e.g., centrally-controlled autonomous vehicles) whose actions indirectly regulate the behavior of other individuals (e.g. human drivers) \citep{wu2018stabilizing}.

In short, the algorithmic and experimental contributions of this paper are:
(1) we propose a meta-gradient approach for the reverse Stackelberg game formulation of incentive design in high-dimensional multi-agent reinforcement learning settings; 
(2) we explain and show experimentally that using standard RL for the incentive designer faces significant difficulties even small finite state finite action Markov games;
(3) our proposed method can converge to known global optima in standard benchmark problems, and it generates significantly higher social welfare than the previous state-of-the-art in a complex high-dimensional economic simulation of market dynamics with taxation.

\vspace{-2pt}
\section{Related work}
\label{sec:related}

A large body of previous work on incentive, utility, and mechanism design belongs to the analytic paradigm, which faces limitations such as linear agent cost and planner incentive functions \citep{ratliff2020adaptive,paccagnan2019utility}, 
finite single-round games \citep{li2018potential},
state-based potential games \citep{li2013designing},
or pertain to special problems such as welfare distribution \citep{gopalakrishnan2014potential} or seller-buyer auctions \citep{shen2019automated}.
These simplifications result in lower applicability to complex, nonlinear, and temporally-extended environments such as dynamic economies \citep{zheng2020ai}.
In contrast, we adopt the paradigm of agent-based simulation \citep{tesfatsion2006handbook} and take state-of-the-art agent learning methods (i.e., deep reinforcement learning, at present) as the starting point to inform a method for incentive design, at the cost of discarding analytical tractability.



Previous work on incentive or mechanism design with RL differ from ours in the choice of the algorithm for the incentive designer or the model of agents.
\citet{brero2020reinforcement,tang2017reinforcement} apply RL to the upper-level planner for non-RL agents.
\citet{pardoe2006adaptive} use perturbation-based gradient ascent to search for hyperparameters of a $k$-armed bandit algorithm that determines the parameters of an auction.
\citet{mguni2019coordinating} employ Bayesian optimization and treat the lower-level multi-agent RL as a black-box.
The central planner in \citet{baumann2020adaptive} optimizes social welfare in 2-player matrix games by anticipating the players' one-step updates.
\citet{li2020end} assume that agents' total payoff is a continuous and differentiable function of the joint strategy---which does not hold in general if agents' original reward can be any combination of discrete and nondifferentiable rules---and differentiate through the  variational inequality reformulation of Nash equilibria.
The closest work to ours are \citet{yang2020learning}, where fully-decentralized RL agents learn mutual pairwise incentivization, and \citet{zheng2020ai,danassis2021achieving}, where a central RL planner optimizes an adaptive tax or price policy at the same time-scale as RL agents' policy optimization.

The technical aspect of our method builds on single-agent meta-gradient RL \citep{xu2018meta} and discovering intrinsic rewards \citep{zheng2018learning}, which we extend to the multi-agent setting and refine with the principle of online cross-validation \citep{sutton1992adapting}.
Related to but different from the variety of existing single-agent meta-learning methods enumerated in \citet[Table~1]{xu2020meta}, our method learns a general neural network representation of an incentive function within a single lifetime, as opposed to methods that optimize hyperparameters \citep{xu2018meta,lorraine2020optimizing,stadie2020learning}, learn target functions \citep{xu2020meta}, or use multiple lifetimes over different environments to find general update functions \citep{kirsch2019improving,oh2020discovering}. 






\vspace{-2pt}
\section{Method}
\label{sec:method}


We propose a method, called ``MetaGrad'', for an Incentive Designer (ID) to optimize a measure of social welfare by explicitly accounting for the impact of incentives on the behavior of a population of $n$ independent agents.
Each agent $i \in \lbrace 1, \dotsc, n \rbrace$ has an individual reward function $R^i \colon \Scal \times \Acal^n \times \Ucal \rightarrow \Rbb$, which depends not only on the global state $s \in \Scal$ and the joint action $\abf \in \Acal^n$, as in standard Markov games \citep{littman1994markov} with transition function $P(s'|s,\abf)$, but also on an incentive that is parameterized by $u \in \Ucal$ for some bounded set $\Ucal \subset \Rbb^l$ (e.g., a vector of marginal tax rates).
We assume that $R^i$ is differentiable with respect to the argument $u$---this holds in the common case of additive incentives such as highway tolling, as well as in complex mechanisms such as a bracketed tax schedule \eqref{eq:tax_formula}.
This incentive $u$ is generated by the ID via a learned incentive function $\mu_{\eta} \colon \Scal \times \Acal^n \rightarrow \Ucal$, parameterized by $\eta$, which adaptively responds to the current system state and joint action of the agents.
Each agent $i$ independently learns a policy $\pi_{\theta^i}$, parameterized by $\theta^i$, to optimize its own individual expected return, while the ID's performance is measured by a social welfare reward $R^{\text{ID}}(s,\abf)$.
Let $\pibf$ and $\thetabf$ denote the agents' joint policy and policy parameters, respectively.
For brevity, we use $R^i_{\eta}(s,\abf)$ to denote $R^i(s,\abf, \mu_{\eta}(s,\abf))$, and we identify $\thetabf$ with the policy $\pibf_{\thetabf}$ where no confusion arises.

The ID aims to solve the bilevel optimization problem
\begin{align}
    &\text{(ID)} \quad \max_{\eta} J^{\text{ID}}(\eta; \widehat{\thetabf}) := \Ebb_{\pibf_{\widehat{\thetabf}}, P(s'|s_t,\abf)} \left[ \sum_{t=0}^H \gamma^t R^{\text{ID}}_t - \psi(s_t) \right] \label{eq:bilevel_objective} \\
    &\text{(Agents) } \widehat{\thetabf} = \left\lbrace \argmax_{\theta^i} J^i(\thetabf; \eta) := \Ebb_{\pibf_{\thetabf}} \left[ \sum_{t=0}^H \gamma^t R^i_{\eta}(s_t,\abf_t) \right] \right\rbrace_{i=1}^n \label{eq:bilevel-constraint}
\end{align}
over episode horizon $H$, and $\psi$ is a known cost for incentivization. 

We assume that the $n$ agents apply a gradient-based reinforcement learning procedure $\text{RL}(\cdot)$ to update their policies in response to the reward determined by $\eta$, i.e., that \eqref{eq:bilevel-constraint} is achieved by $\thetahat^i = \text{RL}(\theta^i_0; \eta)$, where $\theta^i_0$ is an initial policy.
In the ideal case, one should measure the population behavior under the final joint policy $\hat{\thetabf}$, after convergence of the RL process under a fixed $\eta$, to evaluate the effectiveness of $\eta$ in optimizing social welfare and conduct a single $\eta$ update.
However, the high sample count required for convergence of RL in practice is prohibitively expensive, especially if one wishes to apply gradient descent to optimize $\eta$.
To tackle this challenge, we build on the effectiveness of meta-gradient RL in optimizing hyperparameters and parameterized objectives concurrently with an agent's policy optimization \citep{xu2018meta,xu2020meta,oh2020discovering}.
We apply iterative gradient descent to the upper objective \eqref{eq:bilevel_objective} on the same timescale as the agents' policy optimization \eqref{eq:bilevel-constraint}, by explicit differentiating through the agents' policy updates.
We emphasize that even though the ID does not wait for convergence of the final $\hat{\thetabf}$, we follow the principle of online cross-validation \citep{sutton1992adapting} and extend it to the optimal control setting
: the data used for the $\eta$-update is still generated by the agents' {\it updated} policies, not by any arbitrary policy, in order to measure accurately the impact of $\eta$ on the ID's objective through the agents' learning process.
This differs from previous single-agent meta-gradient RL \citep{xu2018meta,xu2020meta}, where the trajectories used for the outer update were not generated by the updated agent policy.

Specifically, we implement the following algorithm (\Cref{alg:m1-pipeline}).
Given the current policy $\thetabf_0$ and incentive function $\mu_{\eta}$, agents collect trajectories $\lbrace \lbrace \tau^i_j \rbrace_{i=1}^n \rbrace_{j=0}^{M-1} \sim \pibf_{\thetabf_0}$ (\Cref{alg:m1-pipeline}, line 3) and conduct $M$ policy update steps (\Cref{alg:m1-pipeline}, line 5): 
\vspace{-5pt}
\begin{align}\label{eq:agent_update}
    \hat{\thetabf}(\eta) \coloneqq \thetabf_{M} = \thetabf_0 + \sum_{j=0}^{M-1} \Delta \thetabf_j(\eta) \, .
    \vspace{-5pt}
\end{align}
Each agent's update $\Delta \theta^i_j(\eta) \propto \nabla_{\theta^i_j} J^i(\theta^i_j;\eta,\tau^i_j)$ depends on the fixed $\eta$.
For example, if agents learn by policy gradient methods, we have $\Delta \theta^i_j \propto \sum_{t=0}^T \nabla_{\theta^i_j} \log \pi_{\theta^i_j}(a^i_t|o^i_t) A^i_t(\tau^i_j;\eta)$, where $A^i_t$ is an advantage function that depends on $\eta$ via $R^i_{\eta}(s,\abf)$.
Now let $\hat{\taubf}$ denote the subsequent trajectory generated by the agents' updated policies $\hat{\thetabf}$ (\Cref{alg:m1-pipeline}, line 6), which serves as the \textit{validation trajectory} that measures the indirect impact of $\eta$ on the ID's return through the agents' learning.
The ID computes and ascends the gradient of objective \eqref{eq:bilevel_objective} w.r.t. $\eta$ via the chain rule (\Cref{alg:m1-pipeline}, line 7)
\begin{align}\label{eq:chain_rule}
\vspace{-5pt}
    \nabla_{\eta} J^{\text{ID}}(\eta; \hat{\thetabf}, \hat{\taubf}) = \sum_{i=1}^n \left( \nabla_{\eta} \hat{\theta}^i(\eta) \right)^{\top} \left( \nabla_{\hat{\theta}^i} J^{\text{ID}}(\eta;\hat{\thetabf},\hat{\taubf}) \right) - \nabla_{\eta} \psi(\tau)
    \vspace{-5pt}
\end{align}
where $\nabla_{\eta} \hat{\theta}^i$ is computed using a replica of the $\theta^i$ update step.



\begin{algorithm}[t]
\caption{Meta-Gradient Incentive Design with pipelining}
\label{alg:m1-pipeline}
\begin{algorithmic}[1]
\Procedure{}{}
\State Initialize all agents' policy parameters $\theta^i$, incentive function parameters $\eta$ 
\State Generate trajectory $\taubf$ using $\thetabf$ and $\eta$
\For{each iteration}
    \State For all agents, update $\hat{\theta}^i$ with $\taubf$ using \eqref{eq:agent_update}
    \State Generate a new trajectory $\hat{\taubf}$ using new $\hat{\thetabf}$ 
    \State Update $\hat{\eta}$ by gradient ascent along \eqref{eq:ppo_gradient} using $\taubf$ and $\hat{\taubf}$
    \State $\taubf \leftarrow \hat{\taubf}$, $\eta \leftarrow \hat{\eta}$, $\theta^i \leftarrow \hat{\theta}^i$ for all $i \in [n]$.
\EndFor
\EndProcedure
\end{algorithmic}
\end{algorithm}

\textbf{Proximal meta-gradient optimization.}
Instead of computing both factors of \eqref{eq:chain_rule}, we can view \eqref{eq:bilevel_objective} as a standard objective in policy-based RL, with the difference that we optimize with respect to $\eta$ instead of the policy parameters $\hat{\thetabf}$.
Hence, one can apply the policy gradient algorithm \citep{sutton1999policy} by replacing $\nabla_{\hat{\thetabf}}$ with $\nabla_{\eta}$, as shown in \citet[Appendix C]{yang2020learning} and used implicitly by \citet{xu2018meta,xu2020meta}.
We extend this viewpoint by showing (in \Cref{subsec:app_ppo}) that trust-region arguments \citep{schulman2015trust} hold for meta-gradients, which justifies the use of a proximal policy optimization (PPO)-type gradient \citep{schulman2017proximal} for the outer optimization:
\begin{equation}\label{eq:ppo_gradient}
\begin{aligned}
    \nabla_{\eta} J^{\text{ID}}(\eta; \hat{\thetabf}, \hat{\taubf}) = \Ebb_{\abf_t, s_t \sim \pibf_{\hat{\thetabf}(\eta)}} &\left[ \min\left( r(\hat{\thetabf};\eta)A_t, \right. \right. \\
    & \left. \left. \text{clip}(r(\hat{\thetabf};\eta), 1-\epsilon, 1+\epsilon)A_t \right) \right] 
\end{aligned}
\end{equation}
\begin{align}\label{eq:ppo_ratio}
    r(\hat{\thetabf};\eta) &\defeq \frac{\nabla_{\eta} \pibf_{\hat{\thetabf}(\eta)}(\abf|s)}{\pibf_{\hat{\thetabf}(\eta)} (\abf|s)} \, ,  
\end{align}
where $A_t \defeq \sum_{l=t}^{T-1} (\gamma \lambda)^{l-1} \delta_l$ is a generalized advantage estimator computed using $\delta_t \defeq R^{\text{ID}}(s_t,\abf_t) + \gamma V(s_{t+1}) - V(s_t)$, critic $V$ for the ID, discount $\gamma$ and $\lambda$-returns.


\subsection{Technical relation to prior multi-agent learning methods for incentivization}
The ``AI Economist'' \citep{zheng2020ai} treats agents' learning as a black-box: it applies standard RL to a central planner who learns an adaptive tax policy concurrently with the agents' policy learning within a fully-decentralized multi-agent economy.
Rather than addressing the bi-level optimization problem, this expands the multi-agent system and exacerbates the already existing problem of non-stationarity in decentralized MARL, which required heuristics such as curriculum learning and tax annealing that are difficult to tune.
In contrast, our method to train the incentive function fundamentally differs from standard RL: the gradient \eqref{eq:ppo_gradient} is taken with respect to the $\eta$ variables of the incentive function, \textit{through the policy updates} $\hat{\thetabf} \leftarrow \thetabf + \Delta \thetabf$ of all the regular RL agents, where $\thetahat^i$ preserves the dependence of each agent's update on $\eta$ in the computational graph.

Our technical method is the centralized analogue of the fully-decentralized pairwise incentivization in LIO \citep{yang2020learning}.
That work begins with the premise that all, or some, agents in the environment are equipped with the LIO learning mechanism, but this may not hold in general environments where no principal opts to use a LIO agent.
In contrast, our work only assumes that agents learn from reward functions that depend on and can be differentiated with respect to incentives, which pertains to more general potential applications.


\section{Experimental setup}
\label{sec:experimental}

We evaluated our approach in three environments: 1) \textit{Escape Room} (ER) \citep{yang2020learning}, a small but deceptively hard pedagogical example that accentuates the core challenges of incentive design for RL agents; 
2) \textit{Cleanup} \citep{hughes2018inequity}, a high-dimensional instance of a sequential social dilemma; 
and 3) the \textit{Gather-Trade-Build} simulation of a market economy with taxation, trading, and competition for limited resource.
We conducted eight independent runs per method for \textit{Escape Room}, 10 for \textit{Cleanup}, and four for \textit{Gather-Trade-Build}.
\Cref{subsec:environments} summarizes the high-level features of these environments; \Cref{app:environment} provides complete specifications.
\Cref{subsec:implementation} describes the implementation of the method and baselines.

\subsection{Environments}
\label{subsec:environments}

\textbf{Escape Room} \citep{yang2020learning}. 
The \textit{Escape Room} game $\text{ER}(n,m)$ is a discrete $n$-player Markov game with individual extrinsic rewards and parameter $m < n$.
An agent gets $+10$ extrinsic reward for exiting a door, but this requires $m$ other agents to sacrifice their own self-interest by pulling a lever at a cost of $-1$ each.
We fix all agents to be standard independent RL agents without ``give-reward'' capabilities, and we introduce a central incentive designer who can modify each agent's reward by adding a scalar bounded in $(0,2)$, since an incentive value of $1 + \epsilon$ for $\epsilon > 0$ is sufficient for an agent to overcome the $-1$ penalty for pulling the lever.
The ID's reward $R^{\text{ID}}$ is the sum of all agents' rewards, and the cost of incentivization $\psi$ is the sum of all scalar incentives.
Hence, the global optimum reward for the incentive designer is $10(n-m) - m - m(1+\epsilon)$.

\textbf{Cleanup} \citep{hughes2018inequity}.
The \textit{Cleanup} scenario is a high-dimensional gridworld sequential social dilemma that serves as a difficult benchmark for independent learning agents.
Agents get +1 individual reward by collecting ``apples'', whose spawn rate decreases in proportion to the amount of ``waste''.
Each agent can contribute to the public good by firing a cleaning beam to clear waste, but doing so would enable other agents to defect and selfishly collect apples, hence posing a difficult social dilemma.
We extended the open-source implementation \citep{vinitsky2020} to provide a global observation image for the incentive designer.

\textbf{Gather-Trade-Build (GTB)} \citep{zheng2020ai}.
The GTB simulation is a 2D grid world in which agents with varying skill levels collect resources that replenish stochastically, spend resources to build houses for coins (i.e., income), and trade coins for resources in an auction system, at the expense of labor costs for each action.
It is not known to be a sequential social dilemma.
GTB has a much larger state and action space than \textit{Escape Room} and \textit{Cleanup} due to the auction system, which provides 44 trading actions and supplements each agent's spatial observation with the current and historical market information (e.g., counts of bids and asks for various price levels for each resource).
At time $t$, the system productivity $\text{prod}_t$ is defined as the sum of all agents' coins, and equality $\text{eq}_t$ is defined such that $\text{eq}=1$ corresponds to uniform coin over agents and $\text{eq}=0$ means only one agent has non-zero coin.
The incentive designer is a central tax planner who optimizes a trade-off between productivity and equality, defined as $\sum_{t=1}^H \text{prod}_t \cdot \text{eq}_t$ over horizon $H$, by imposing taxes according the following mechanism.
Each episode lasts for $H=100$ time steps and consists of $10$ tax periods.
At the start of each tax period, the ID sets a tax schedule $T \colon \Rbb \rightarrow \Rbb$ that determines the tax $T(z)$ applied to an agent's income $z$ earned within the period.
$T$ is a bracketed tax schedule based on the US federal taxation scheme: given a set of income thresholds $\lbrace m_b \rbrace_{b=0}^{B=7}$ with $m_0 = 0 < m_1 < \dotsm < m_B = \infty$, the ID defines $T$ by setting a vector of \textit{marginal tax rates} $[\tau_b]_{b=1}^B$, where $\tau_b \in [0,1]$ applies to bracket $(m_b, m_{b+1})$, such that the total tax on income $z$ is given by
\begin{align}\label{eq:tax_formula}
    T(z) = \sum_{b=0}^{B-1} \tau_b \left( (m_{b+1} - m_b) \mathbf{1}_{z > m_{b+1}} + (z-m_b) \mathbf{1}_{m_b < z \leq m_{b+1}} \right)
\end{align}
where $\mathbf{1}_p = 1$ if $p$ is true and 0 otherwise.
At the end of each tax period, the total collected tax is evenly distributed back to all agents:
if agent $i$ gets total income $z^p_i$ within period $p$, then the agent's final adjusted income at the end of the tax period is given by:
\begin{align}\label{eq:adjusted-income}
    \tilde{z}^p_i = z^p_i - T(z^p_i) + \frac{1}{N} \sum_{j=1}^n T(z^p_j) \, .
\end{align}
We used the $15\times 15$ map called ``env-pure\_and\_mixed-15x15'' \citep{zheng2020ai}, which features similar spatial distribution of resource spawn points as the original $25\times 25$ map.
Each agent's observation consists of an $11\times 11$ egocentric spatial window, their resource inventories, collection and building skills, personal and other agents' bids and asks, and quantities derived from the current period's tax rate.
The ID observes the complete spatial world state, agents' inventories and incomes, cumulative bids and asks, and all derived tax quantities, but does not know agents' private skill and utility functions.
Agents have the same discrete action space consisting of movements, building, and trading actions.
\Cref{app:foundation} provides more information on observation/action spaces and agent utilities.

\begin{figure*}[t]
\centering
\begin{subfigure}[t]{0.25\linewidth}
    \centering
    \includegraphics[width=1.0\linewidth]{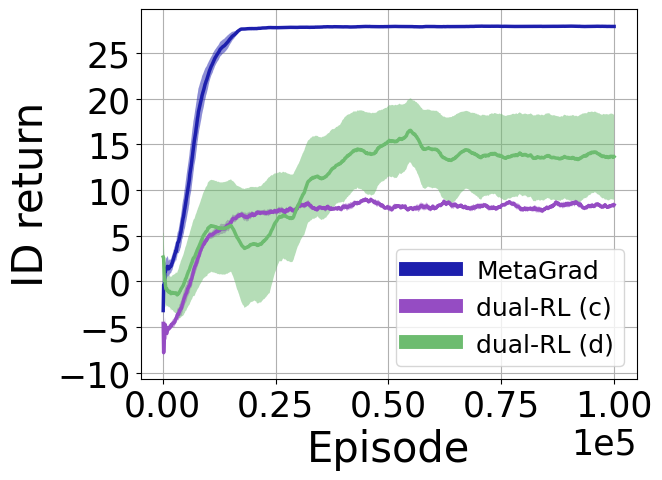}
    \caption{ER(5,2)}
    \label{fig:er_n5m2}
\end{subfigure}
\hfill
\begin{subfigure}[t]{0.25\linewidth}
    \centering
    \includegraphics[width=1.0\linewidth]{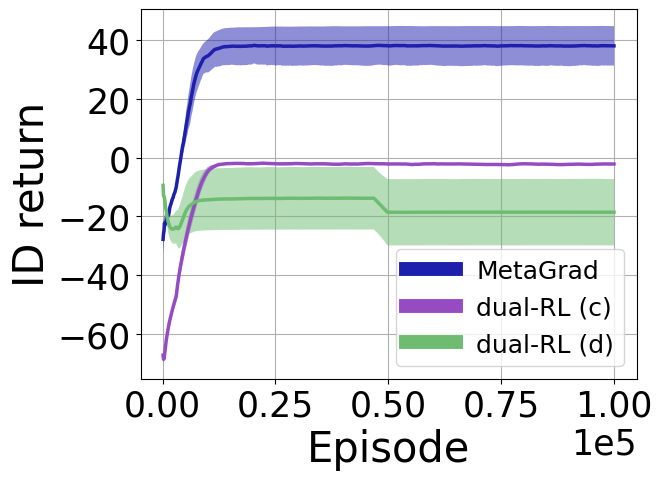}
    \caption{ER(10,5)}
    \label{fig:er_n10m5}
\end{subfigure}
\hfill
\begin{subfigure}[t]{0.23\linewidth}
    \centering
    \includegraphics[width=1.0\linewidth]{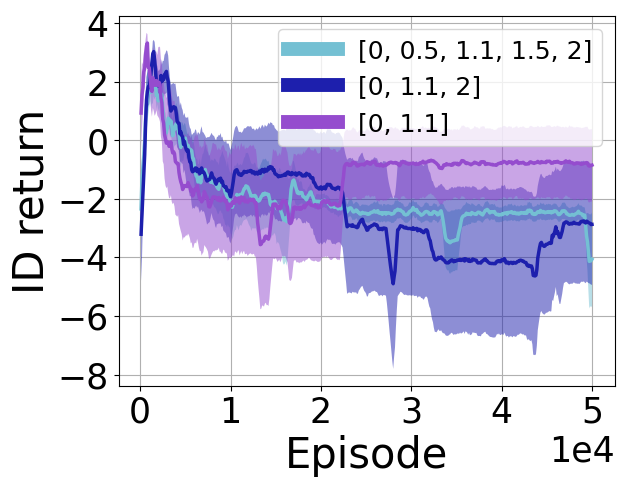}
    \caption{Dual-RL (d)}
    \label{fig:n2m1_dual_d}
\end{subfigure}
\hfill
\begin{subfigure}[t]{0.25\linewidth}
    \centering
    \includegraphics[width=1.0\linewidth]{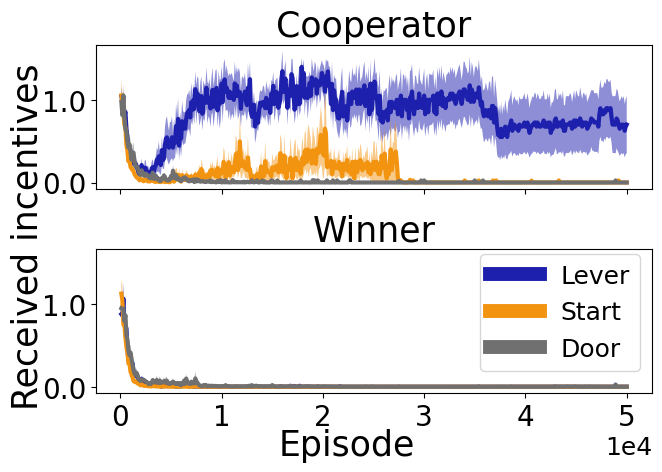}
    \caption{Incentives}
    \label{fig:n2m1_received}
\end{subfigure}
\caption{Escape Room. (a) In $\text{ER}(5,2)$, MetaGrad converges to the global optimum where two agents are incentivized by $1+\epsilon$ (for some small $\epsilon > 0$) to pull the lever with -1 penalty, three other agents exit the door with +10 reward, which results in a total ID reward near 26. (b) MetaGrad also converges to the global optimum in $\text{ER}(5,2)$ with ID reward near 40.
(c) Dual-RL (discrete) was unstable and had high variance across seeds, for various choices of discrete action spaces for the incentive values.
(d) Incentives given by MetaGrad for each action by each agent in $\text{ER}(2,1)$.
}
\label{fig:results_er}
\end{figure*}

\subsection{Implementation and baselines}
\label{subsec:implementation}

We describe the key implementation of all methods here and include all remaining details in \Cref{subsec:app_implementation}.
We use $M=1$ for MetaGrad across all experiments, such that an ID update occurs after each policy update by agents.
We employ pipelining to improve the efficiency of MetaGrad: the validation trajectory $\hat{\taubf}$ generated by agents' updated policies (\Cref{alg:m1-pipeline}, line 6), which is required for the ID's update step, is used for the agents' policy update in the next iteration (\Cref{alg:m1-pipeline}, line 5).
To differentiate through the agents' learning step, MetaGrad has access to agents' policy parameters and gradients.
This assumption can be removed by using behavioral cloning to obtain surrogate models of agents, which has been demonstrated by existing methods that rely on knowledge of agent parameters \citep{foerster2018learning,yang2020learning}.

Our main baseline is termed \textbf{dual-RL}, in which the incentive designer itself is a standard RL agent who optimizes the system-level objective at the same time-scale as the original RL agents.
Dual-RL is the centralized analogue of the decentralized agents with ``give-reward'' actions, introduced as a baseline in \citet{yang2020learning}.
It is also formally equivalent to the method called ``AI economist'' in \citet{zheng2020ai}.
In \textit{Escape Room} , we compare with discrete-action and continuous-action variants of dual-RL, labeled ``dual-RL (d)'' and ``dual-RL (c)''.
In \textit{Cleanup}, we compare with ``dual-RL (c)''.
In GTB, we implemented the core aspects of the ``AI Economist'' based on available information in \citet{zheng2020ai} (relabeled as ``dual-RL'' here), and also compare with the static US federal tax rates.
We tuned hyperparameters for all methods equally using a successive elimination method, detailed in \Cref{subsec:app_hyperparameters}.

\textbf{Escape Room.}
We used policy gradient \citep{sutton1999policy} without parameter sharing as the base agent implementation for all methods.
The incentive function $\mu_{\eta}$ in MetaGrad is a neural network that maps the global state and agents' joint action to a scaled sigmoid output layer of size $|\Acal|=3$ (the number of possible agent actions), such that the value of each output node $i$ lies in $(0, 2)$ and is interpreted as the incentive for action $i$ taken by any agent.
This parameterization enables MetaGrad to scale to larger number of agents, e.g. ER$(10,5)$.
The cost for incentivization is the sum of all incentives given to agents, and is accounted by $\psi_t(\eta)$ in MetaGrad's loss function.\footnote{At \textit{train time}, one cannot for cost in $R^{\text{ID}}$ of the current episode because MetaGrad only learns from $R^{\text{ID}}$ in the next episode, not the episode where incentives are given. At \text{test time}, incentives are subtracted from $R^{\text{ID}}$ so that comparison to baselines is fair.}
For \textbf{dual-RL (d)}, we tried three different sets of discrete incentives $S_r = \lbrace 0, 1.1 \rbrace$, $S_r =\lbrace 0, 1.1, 2.0 \rbrace$, and $S_r =\lbrace 0, 0.5, 1.0, 1.5, 2.0 \rbrace$.
The designer's action space is Discrete($|S_r|^{|\Acal|}$).
Hence, the designer's action is an assignment of a scalar incentive value to each possible agent action, and the policy output is a categorical distribution.
In \textbf{dual-RL (c)}, the designer's action space is $(0,2)^n$, and its policy $\pi(a_{\text{ID}}|o)$ is defined by sampling $u \sim \Ncal(\mu_{\eta}(s,\abf), \mathbf{1})$ with neural network $\mu_{\eta} \colon \Scal \times \Acal^n \rightarrow \Rbb^n$, then applying the same sigmoid output layer $\sigma$ as MetaGrad to get $a_{\text{ID}} = \sigma(u)$.
The total incentives given to agents are subtracted from $R^{\text{ID}}$.
To compare with the method of \citet{baumann2020adaptive}, we implemented separate experiments with actor-critic for agents' policy updates.

\textbf{Cleanup.}
We used actor-critic agents with TD(0) critic updates \citep{sutton2018reinforcement} and parameter-sharing as the base agent for all methods.
MetaGrad and dual-RL (c) have the same architecture as for Escape Room, except that: 1) the observation input for agents and the ID is an RGB image; 2) the ID has a vector observation indicating whether or not each agent performed a cleaning action; 3) each output node of the incentive function is interpreted as the incentive for an action type in the set $\lbrace \text{fire cleaning beam, collect apples, else} \rbrace$.

\textbf{Gather-Trade-Build.}
We used PPO agents \citep{schulman2017proximal} with parameter sharing for all methods.
The incentive function in MetaGrad has $B=7$ output nodes, where the value $\tau_b$ at each node $b$ is capped by sigmoid activation to lie in $(0,1)$ and is interpreted as the tax rate $\tau_b$ for bracket $(m_b,m_{b+1})$.
By \eqref{eq:tax_formula}, \eqref{eq:adjusted-income}, and \eqref{eq:utility}, each agent's policy update is a differentiable function of the incentive function parameters $\eta$.
The ID has seven action subspaces (one for each of the $B=7$ tax brackets), each with 21 discrete actions that choose the marginal tax rate in $\lbrace 0, 0.05,\dotsc, 1.0\rbrace$.
Dual-RL applies standard RL to the ID, whose action space is a direct product of seven action subspaces (one for each of the $B=7$ tax brackets), each with 21 discrete choices of the marginal tax rate in $\lbrace 0, 0.05,\dotsc, 1.0\rbrace$.
\citet{zheng2020ai} reported the need for a two-phase curriculum with tax annealing to stabilize training for Dual-RL, which may face difficulties with non-stationarity of the expanded multi-agent system.
Hence, in the curriculum version of GTB experiments, we first train agents in a free-market (zero tax) scenario in Phase 1, while the RL-based tax policy is introduced in Phase 2 with a gradual annealing schedule on the maximum tax rate.
We repeated experiments without curriculum to investigate the stability of MetaGrad.
For measurements of economic activity and tax rates produced by trained models, we report the mean and standard error over the four trained models of the mean over 100 test episodes per model.

\begin{figure*}[t]
\centering
\begin{minipage}{0.72\linewidth}
\begin{subfigure}[t]{0.32\linewidth}
    \centering
    \includegraphics[width=1.0\linewidth]{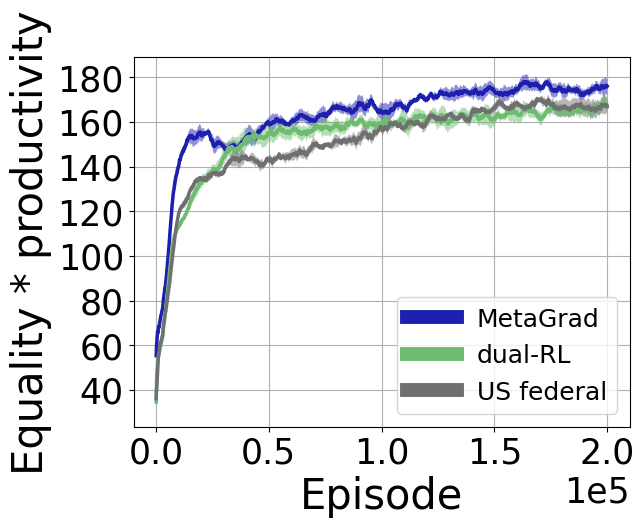}
    \caption{Social welfare}
    \label{fig:15x15_nocurr}
\end{subfigure}
\hfill
\begin{subfigure}[t]{0.32\linewidth}
    \centering
    \includegraphics[width=1.0\linewidth]{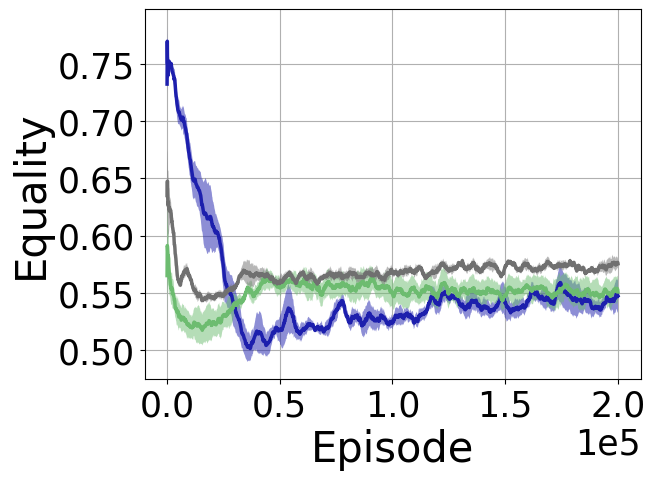}
    \caption{Equality}
    \label{fig:15x15_nocurr_eq}
\end{subfigure}
\hfill
\begin{subfigure}[t]{0.32\linewidth}
    \centering
    \includegraphics[width=1.0\linewidth]{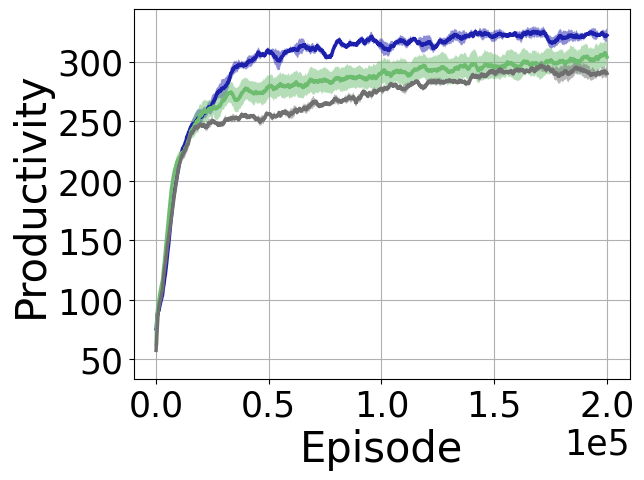}
    \caption{Productivity}
    \label{fig:15x15_nocurr_prod}
\end{subfigure}
\caption{GTB without curriculum: MetaGrad finds tax policies that induce higher social welfare than baselines, by promoting higher productivity at similar levels of equality.}
\label{fig:foundation_nocurr}
\end{minipage}
\hfill
\begin{minipage}{0.25\linewidth}
    \centering
    \includegraphics[width=1.0\linewidth]{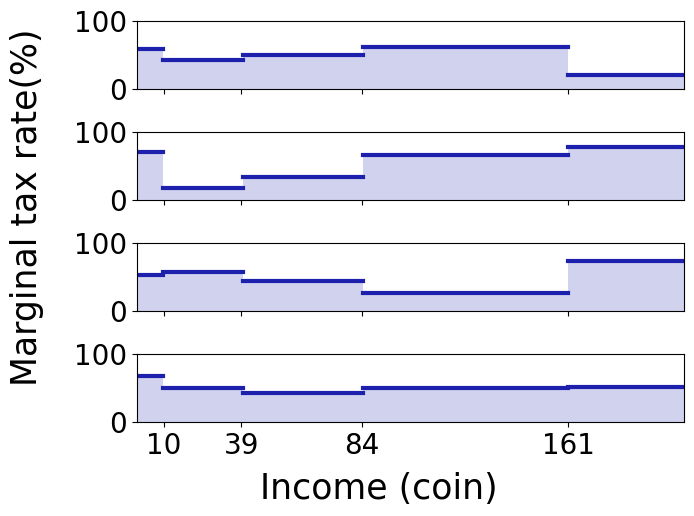}
    \vspace{2pt}
    \caption{MetaGrad tax rates for each independent run.}
    \label{fig:15x15_nocurr_tax_rate_m1}
\end{minipage}
\end{figure*}

\begin{figure*}[t]
\centering
\begin{minipage}{0.72\linewidth}
\begin{subfigure}[t]{0.32\linewidth}
    \centering
    \includegraphics[width=1.0\linewidth]{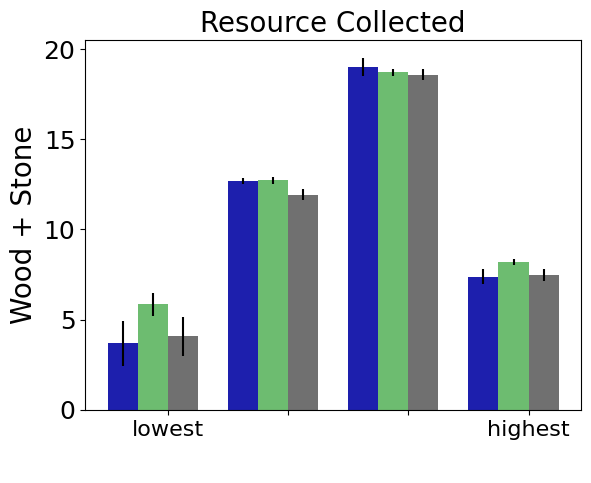}
    \caption{Resource collected}
    \label{fig:15x15_nocurr_resource}
\end{subfigure}
\hfill
\begin{subfigure}[t]{0.32\linewidth}
    \centering
    \includegraphics[width=1.0\linewidth]{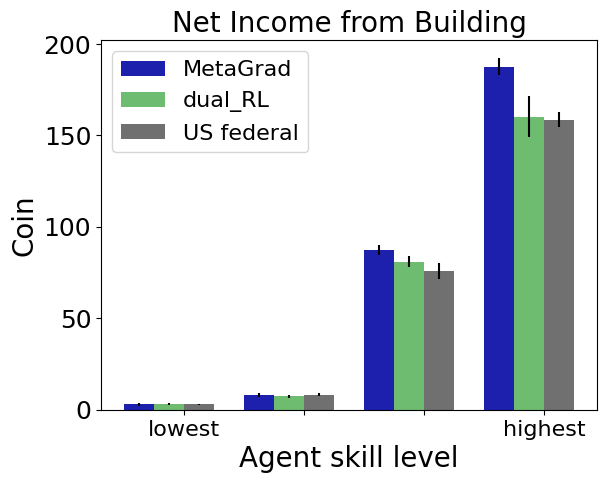}
    \caption{Income from building}
    \label{fig:15x15_nocurr_income_build}
\end{subfigure}
\hfill
\begin{subfigure}[t]{0.32\linewidth}
    \centering
    \includegraphics[width=1.0\linewidth]{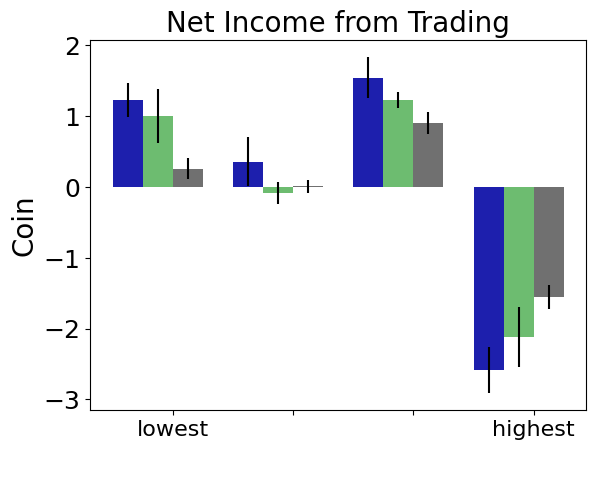}
    \caption{Income from trading}
    \label{fig:15x15_nocurr_income_trade}
\end{subfigure}
\caption{GTB without curriculum: economic activity after 200k training episodes.}
\label{fig:15x15_nocurr_activity}
\end{minipage}
\hfill
\begin{minipage}{0.25\linewidth}
    \centering
    \includegraphics[width=1.0\linewidth]{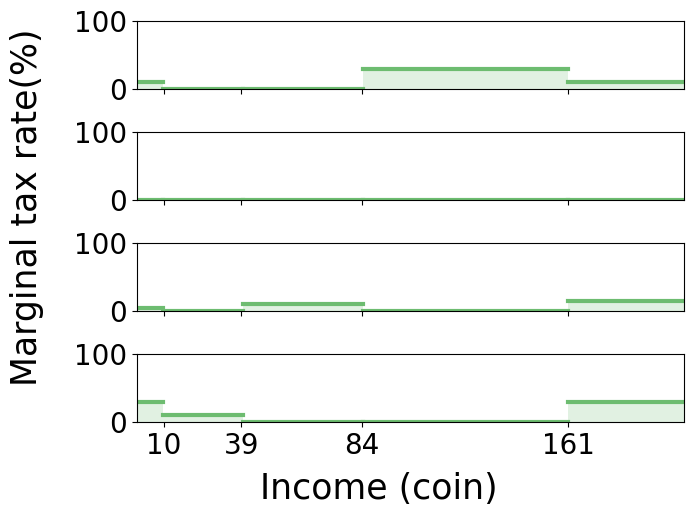}
    \vspace{1pt}
    \caption{Dual-RL tax rates for each independent run.}
    \label{fig:15x15_nocurr_tax_rate_dual_RL}
\end{minipage}
\end{figure*}

\section{Results}
\label{sec:results}



\label{subsec:results_er}
\textbf{Escape Room}.
MetaGrad converged to the known global optimum value of approximately 26 in ER$(5,2)$ and 40 in ER$(10,5)$ (\Cref{fig:er_n5m2,fig:er_n10m5}, respectively).
\Cref{fig:n2m1_received} shows the dynamics of incentivization during training in the case of ER$(2,1)$, where we labeled each agent \textit{at the end of training} as either a ``cooperator'' or a ``winner'' based on whether the agent primarily pulls the lever or exits the door, respectively.
We see that the cooperator consistently receives incentives of $1+\epsilon$ during the majority of episodes, which explains the emergence of its cooperative behavior, whereas the winner receives zero incentives asymptotically, which shows the designer learned to avoid unnecessary costs.
In contrast, both dual-RL (d) and dual-RL (c) did not solve ER, even including various choices of the discrete action space (\Cref{fig:n2m1_dual_d}).
This is because a standard RL incentive designer optimizes the expected return of \textit{one} episode, but the impact of its "give-reward" action only appears after agents have conducted learning updates over \textit{many} episodes.
In any given episode, the ID's reward contains no information about the impact of the actions it chose during that episode.
The only conceivable way that dual-RL learns is by serendipity: the ID's action $a$ in a \textit{previous} episode $i$ led to a change in agents' behavior that results in positive reward for the ID during the \textit{current} episode $j$, and the ID happened to take $a$ again in episode $j$, which results in correct credit assignment when learning from episode $j$.
In fact, among eight independent runs for dual-RL, the only run that succeeded had the same random seed as in hyperparameter search.
\Cref{app:additional-results}, \Cref{fig:n2m1_amd}, shows that MetaGrad outperforms the incentive design method of \citet{baumann2020adaptive}, which faced numerical instabilities when extended from matrix games to Markov games.

\setlength{\columnsep}{5pt}
\begin{wrapfigure}{l}{0.23\textwidth}
\vspace{-10pt}
\hspace{-2pt}
    \centering
    \includegraphics[width=1.0\linewidth]{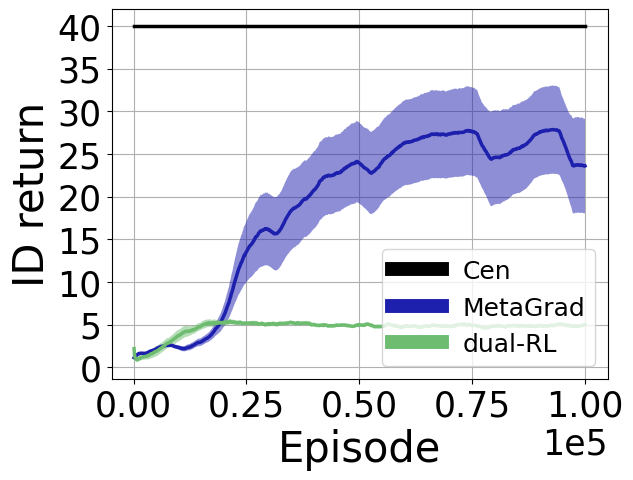}
    \vspace{-20pt}
    \caption{7x7 map}
    \label{fig:ssd_7x7}
    \vspace{-10pt}
\end{wrapfigure}
\textbf{Cleanup}.
As shown in \Cref{fig:ssd_7x7}, MetaGrad achieved a high level of social welfare, which is only possible because one agent received incentives to take cleaning actions while another agent collects apples.
The method labeled ``Cen'' trains a single policy that acts for both agents; it serves as an empirical upper bound on performance.
Under dual-RL, agents did not receive appropriate incentives and hence behaved selfishly---occasionally using the cleaning beam but immediately competing for any apples that spawned---and therefore converged to low social welfare.

\begin{figure*}[t]
\centering
\includegraphics[width=0.24\linewidth]{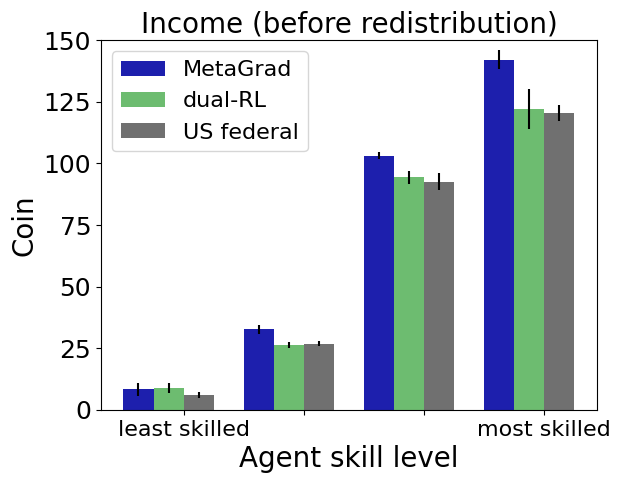}  
\includegraphics[width=0.24\linewidth]{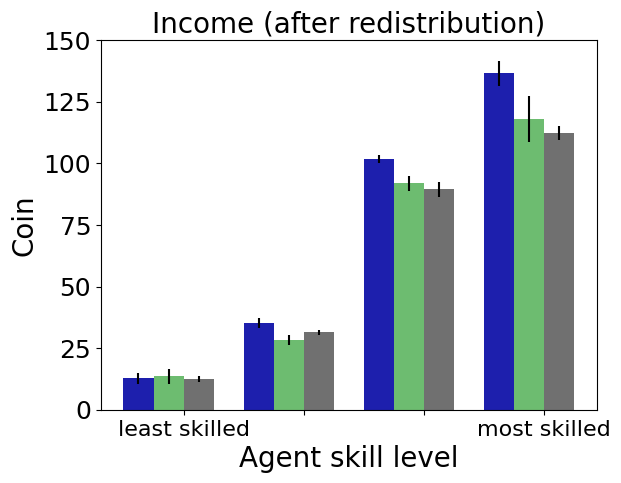}   
\includegraphics[width=0.24\linewidth]{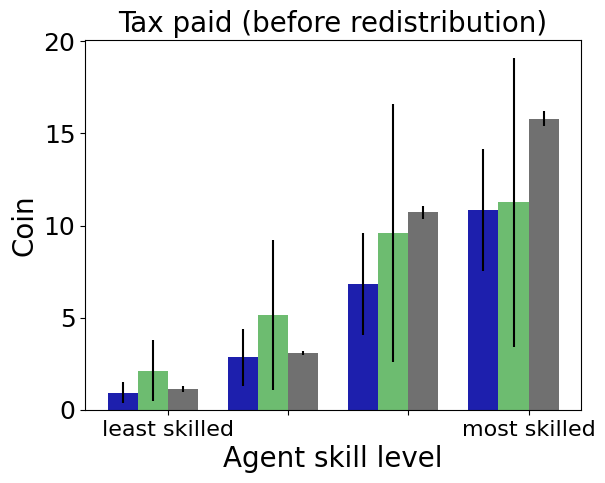}   
\includegraphics[width=0.24\linewidth]{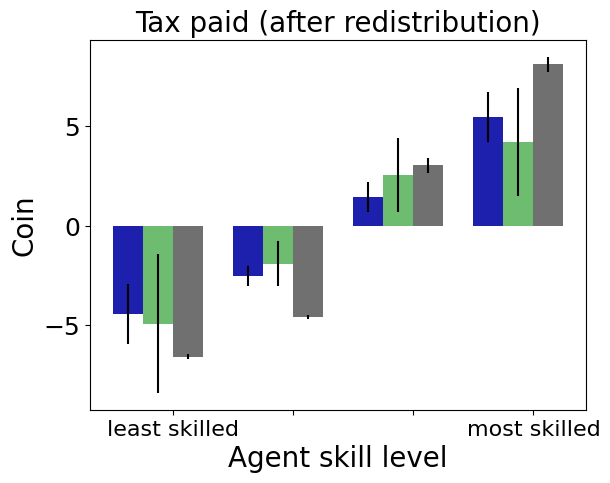}   
\caption{GTB without curriculum: income and tax before and after redistribution, after 200k training episodes.}
\label{fig:15x15_nocurr_redistribution}
\end{figure*}

\begin{figure*}[t]
\centering
\includegraphics[width=0.24\linewidth]{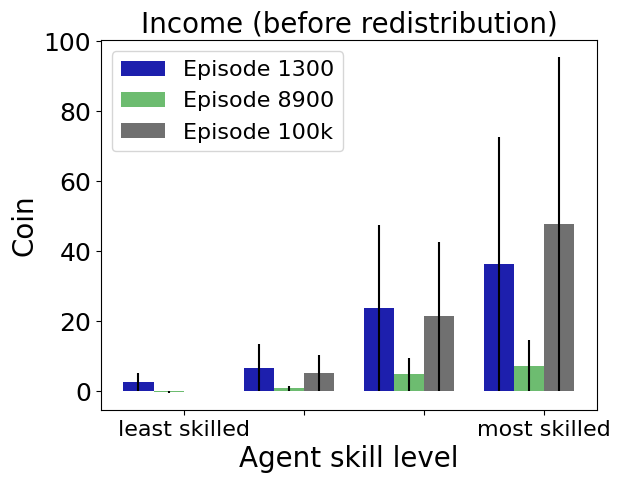}  
\includegraphics[width=0.24\linewidth]{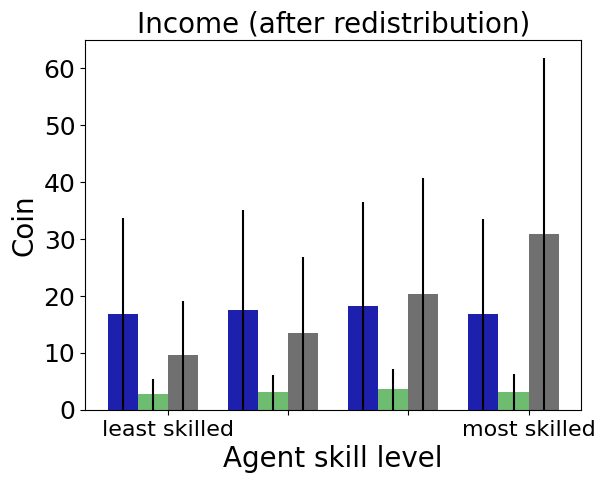}   
\includegraphics[width=0.24\linewidth]{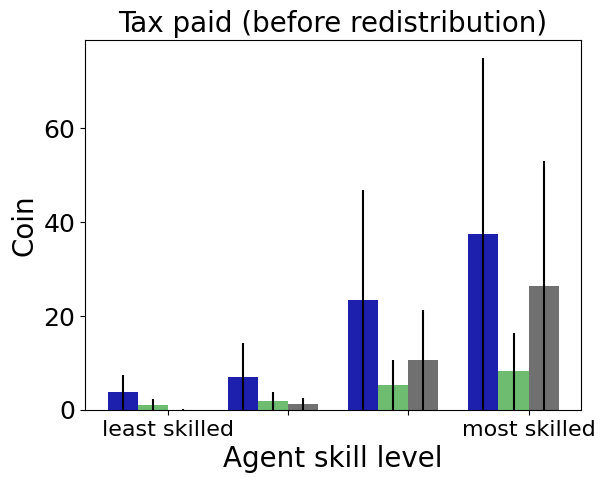}   
\includegraphics[width=0.24\linewidth]{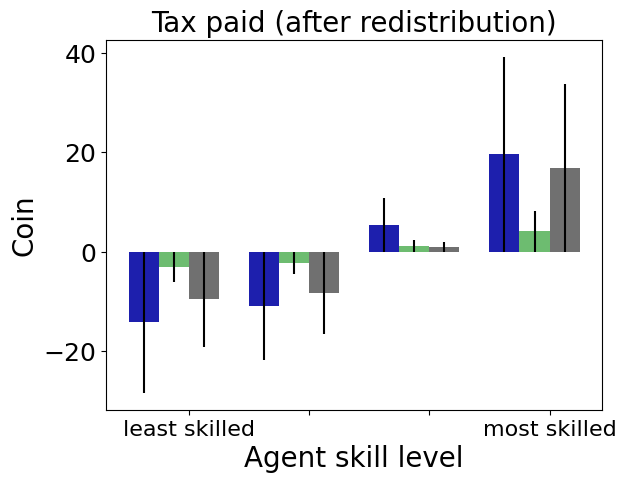}   
\includegraphics[width=0.26\linewidth]{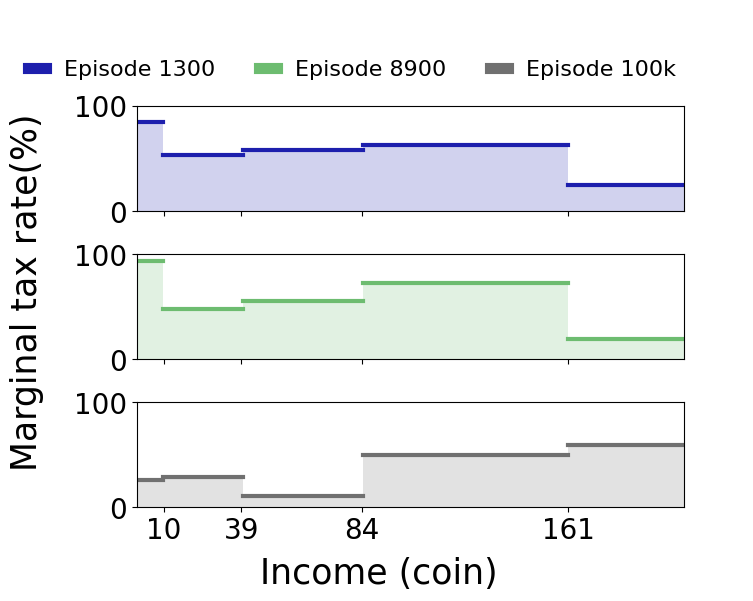}
\includegraphics[width=0.24\linewidth]{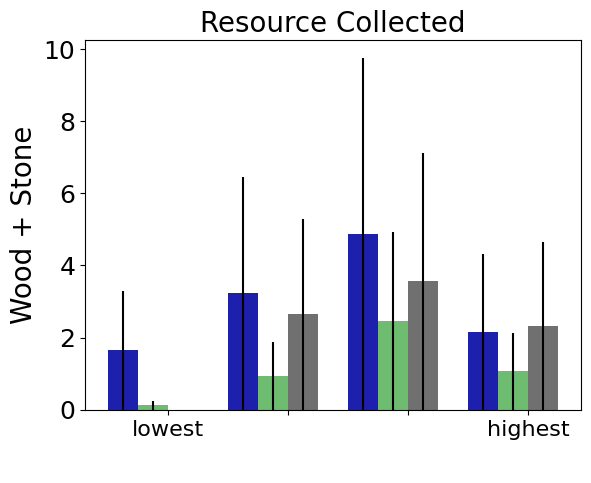}  
\includegraphics[width=0.24\linewidth]{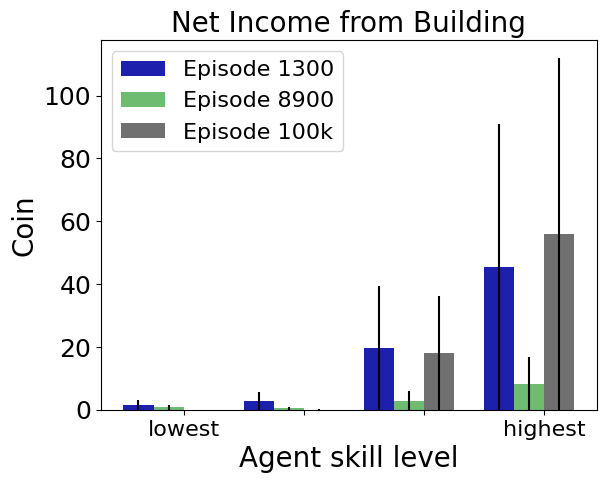}   
\includegraphics[width=0.24\linewidth]{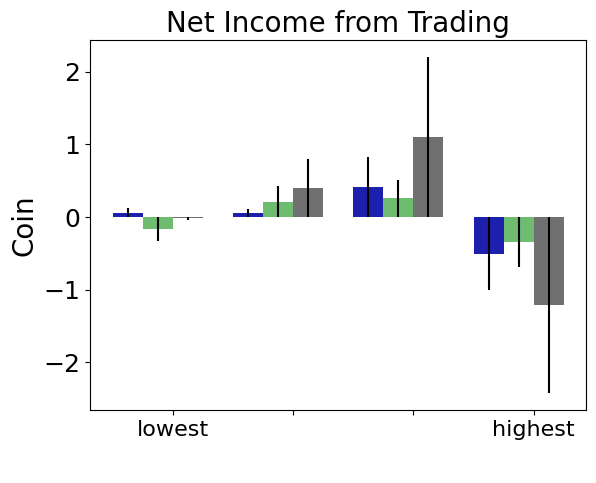}
\vspace{-10pt}
\caption{Training dynamics. Status of agent and designer behavior at 1300, 8900, and 100k episodes. 
Top row: income and tax before and after redistribution. 
Bottom row: tax rates and economic activity.}
\vspace{-10pt}
\label{fig:15x15_phase2_m1_masterbranch_2_dynamics}
\end{figure*}

\vspace{-10pt}
\subsection{Gather-Trade-Build}
\label{subsec:results_foundation}

\begin{wrapfigure}{r}{0.18\textwidth}
    \vspace{-15pt}
    \includegraphics[width=1.0\linewidth]{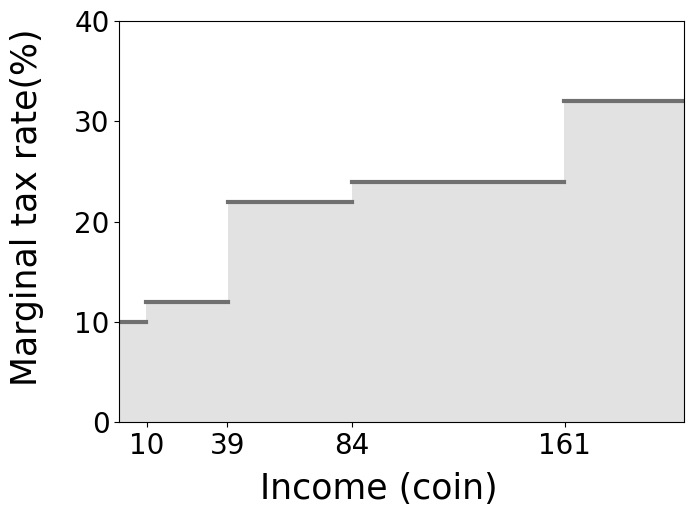}
    \vspace{-20pt}
    \caption{US federal}
    \label{fig:tax_rate_us_federal}
    \vspace{-13pt}
\end{wrapfigure}

\textbf{Social welfare}.
MetaGrad outperforms both dual-RL and US federal without requiring heuristics such as curriculum learning and tax annealing (\Cref{fig:15x15_nocurr}).
It discovers tax rates which differ from the static US federal tax rates in two notable aspects (compare \Cref{fig:15x15_nocurr_tax_rate_m1} and \Cref{fig:tax_rate_us_federal}).
Firstly, MetaGrad imposes much higher taxes than US federal on the lowest income bracket (e.g., 0-10 coin), but chooses relatively lower taxes for the next income bracket (10-39).
Hence, compared to US federal, there is less incentive for agents to fall in the lowest bracket, which may explain the higher income of agents under MetaGrad versus US federal (\Cref{fig:15x15_nocurr_redistribution}).
Secondly, the highest income bracket does not face significantly higher tax rates than other brackets, and even gets the lowest rate in one instance.
While this results in lower equality than US federal (\Cref{fig:15x15_nocurr_eq}), the increase in economic activity---such as resource collection, building, and trading (\Cref{fig:15x15_nocurr_activity})---improves system productivity (\Cref{fig:15x15_nocurr_prod}) and ultimately produces significantly higher social welfare.

\begin{figure*}[t]
\centering
\begin{minipage}{0.72\linewidth}
\begin{subfigure}[t]{0.32\linewidth}
    \centering
    \includegraphics[width=1.0\linewidth]{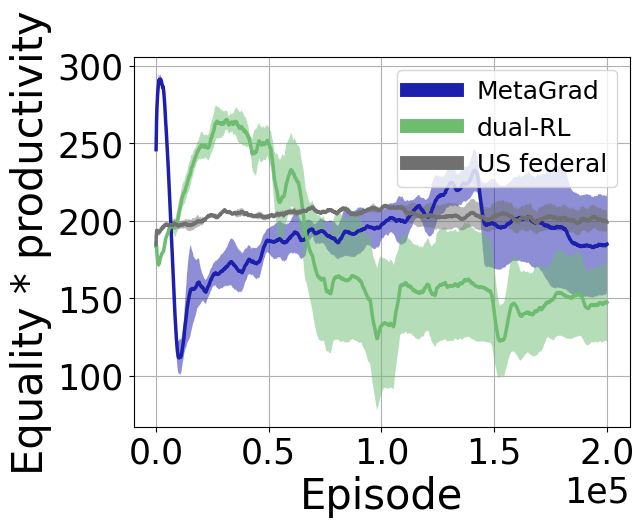}
    \caption{Social welfare}
    \label{fig:15x15_phase2}
\end{subfigure}
\hfill
\begin{subfigure}[t]{0.32\linewidth}
    \centering
    \includegraphics[width=1.0\linewidth]{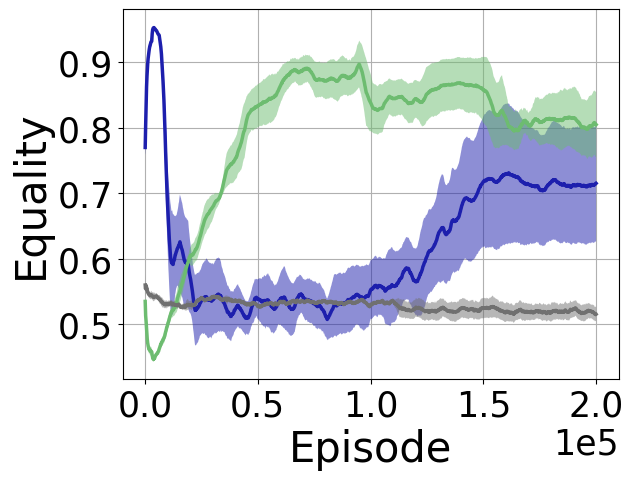}
    \caption{Equality}
    \label{fig:15x15_phase2_eq}
\end{subfigure}
\hfill
\begin{subfigure}[t]{0.32\linewidth}
    \centering
    \includegraphics[width=1.0\linewidth]{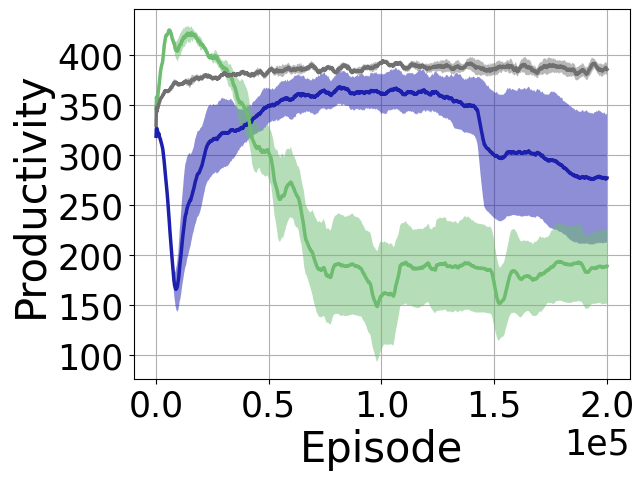}
    \caption{Productivity}
    \label{fig:15x15_phase2_prod}
\end{subfigure}
\vspace{-5pt}
\caption{GTB with curriculum. Both MetaGrad and dual-RL find transient states of high social welfare but are less stable than US federal.}
\label{fig:results_foundation_ppo}
\end{minipage}
\hfill
\begin{minipage}{0.25\linewidth}
    \centering
    \includegraphics[width=1.0\linewidth]{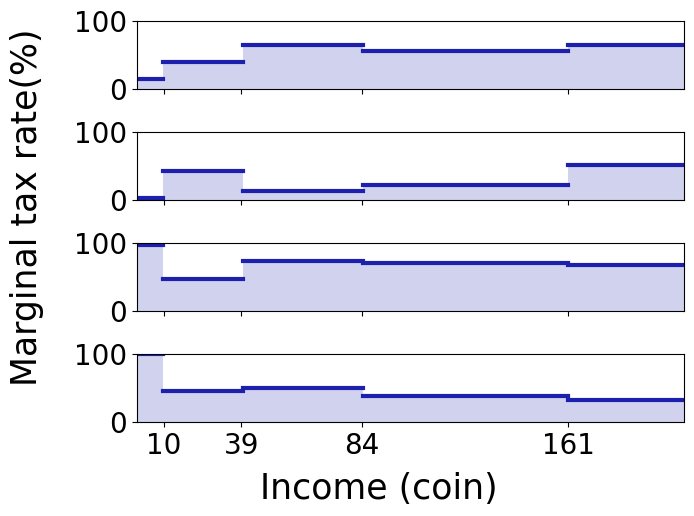}
    \vspace{1pt}
    \caption{MetaGrad tax rates with curriculum.}
    \label{fig:15x15_phase2_tax_rate_m1}
\end{minipage}
\end{figure*}

When curriculum learning is applied to all methods---i.e., for all methods, initializing agents with the same policy that was pre-trained in a free market context---MetaGrad also exceeds the performance of dual-RL.
Except for one out of four runs where social welfare abruptly dropped around 150k episodes, MetaGrad also outperforms US federal.
Similar to the training dynamics observed in \citet{zheng2020ai}, we also observe a transient period where dual-RL passes through unstable local optima (\Cref{fig:15x15_phase2}, near 25k episodes); however, in contrast to their results, dual-RL did not manage to surpass US federal in asymptotic performance.
This is likely because the sudden introduction of a tax planner in Phase 2
may do more harm than good for stability, especially when the extra hyperparameters introduced by curriculum learning and tax annealing are hard to tune.
Both MetaGrad and dual-RL enact higher taxes at lower income brackets than US federal.

\begin{figure*}[t]
\centering
\begin{minipage}{0.72\linewidth}
\begin{subfigure}[t]{0.32\linewidth}
\includegraphics[width=1.0\linewidth]{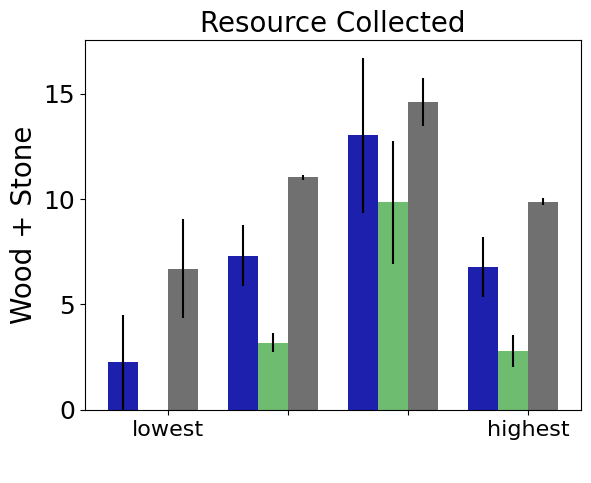}
\caption{Resource collection}
\label{fig:15x15_phase2_resource}
\end{subfigure}
\hfill
\begin{subfigure}[t]{0.32\linewidth}
\includegraphics[width=1.0\linewidth]{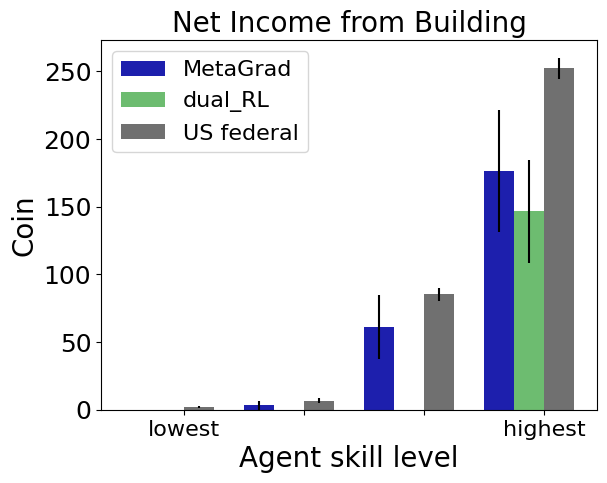}
\caption{Building income}
\end{subfigure}
\hfill
\begin{subfigure}[t]{0.32\linewidth}
\includegraphics[width=1.0\linewidth]{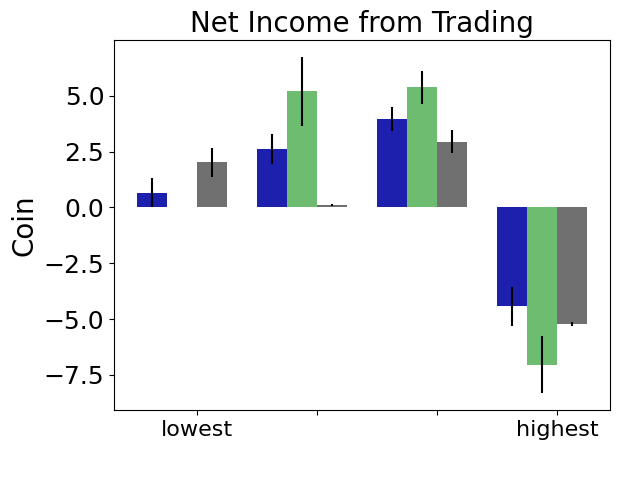}
\caption{Trading income}
\end{subfigure}
\vspace{-5pt}
\caption{GTB with curriculum: economic activity after 200k training episodes in Phase 2.}
\label{fig:activity}
\end{minipage}
\hfill
\begin{minipage}{0.25\linewidth}
    \includegraphics[width=1.0\linewidth]{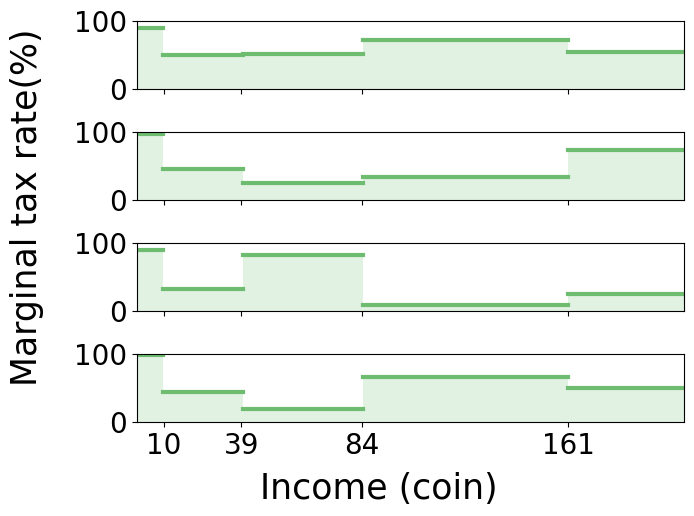}
    \caption{Dual-RL tax with curriculum.}
    \label{fig:15x15_phase2_tax_rate_dual_RL}
\end{minipage}
\vspace{-5pt}
\end{figure*}

\textbf{Economic activity}.
Because skill levels determine the amount of coins per house built, differences in agents' skill levels generate income inequality and behavioral specialization.
For example, agents with second-highest skill tend to collect the most resources and sell them for income, whereas agents with the highest skill spend less effort on resource collection but generate income by building houses from purchased resources (\Cref{fig:15x15_nocurr_resource,fig:15x15_nocurr_income_trade}).
Notably, under all tax policies, all except the highest skill agents receive net positive income from trading (\Cref{fig:15x15_nocurr_income_trade}).
Even though resource collection is comparable across methods, tax policies found by MetaGrad encouraged highest trading activity and hence highest overall income from building, compared to US federal and dual-RL.
In the curriculum case, dual-RL tax policies impose high taxation (above 50\%) on the three lowest income brackets \Cref{fig:15x15_phase2_tax_rate_dual_RL}.
This may explain the fact that the lowest-skilled agent collects zero resources (\Cref{fig:15x15_phase2_resource}), which lowers overall system productivity.

\textbf{Taxation and income}.
Agents with the two highest skill levels pay significantly less tax for tax policies found by MetaGrad than they do for the US federal tax rates, whereas agents with the two lowest skill pay comparably equal tax (\Cref{fig:15x15_nocurr_redistribution}).
This means that MetaGrad does better than US federal at encouraging higher skilled agents to increase economic activity such as building and trading, without affecting resource collection by lower skill agents (as can be seen in \Cref{fig:15x15_nocurr_resource}).
While this comes at the expense of lower equality (\Cref{fig:15x15_nocurr_eq}), the incomes of lower-skilled agents both before and after redistribution are actually higher for MetaGrad than US federal, because lower-skilled agents benefit from increased trading activity (\Cref{fig:15x15_nocurr_income_trade}).
Dual-RL tax policies caused agents of all skill levels to pay more taxes than MetaGrad, which is correlated in overall lower building and trading activity.

\textbf{Training dynamics}.
In GTB without curriculum (\Cref{fig:15x15_nocurr}), the fact that social welfare rises while equality drops in the early 50k episodes is due to \textit{tabula rasa} learning with heterogeneous skills.
In the curriculum setting, the changing tax rates under MetaGrad produced a more dynamical social welfare curve than the fixed US federal rates during training (\Cref{fig:15x15_phase2}). 
This can be useful for extracting potential causal relations between taxation and agents' economic behavior.
For one run of MetaGrad, we measured taxation, income, and economic activity over 100 test episodes at 1300, 8900, and 100k episodes during training, corresponding to the early peak, valley, and steady rising region in the social welfare curve in \Cref{fig:15x15_phase2}.
These measurements are shown in \Cref{fig:15x15_phase2_m1_masterbranch_2_dynamics}, and we make the following observations.
Social welfare is highest at episode 1300, but agents actually have lower income (before redistribution) at episode 1300 than at episode 100k. 
Because taxes do not have instantaneous effect on productivity, MetaGrad quickly learned that social welfare can be artificially inflated by finding a tax scheme to increase the equality index---hence the sharp early peak in \Cref{fig:15x15_phase2_eq}.
As shown in \Cref{fig:15x15_phase2_m1_masterbranch_2_dynamics}, MetaGrad's tax rates at episode 1300 produced nearly uniform income (after redistribution) over all skill levels, by enacting high taxes on the higher skilled agents.
This tax policy disincentivizes agents from earning high income, since episode 1300 is followed by a precipitous drop in productivity (\Cref{fig:15x15_phase2_prod}) that reaches a global minimum near episode 8900, where agent activity levels are lowest (\Cref{fig:15x15_phase2_m1_masterbranch_2_dynamics}).
Nonetheless, MetaGrad adapted its tax policy to produce a recovery of productivity at relatively constant equality, from episodes 25k to 100k (\Cref{fig:15x15_phase2_eq,fig:15x15_phase2_prod}).
The tax policy at episode 100k resembles the progressive schedule of the US federal policy, albeit with significantly lower rates for the 39-84 income bracket that applies to the high-skilled agents except for the very top earners.

\textbf{Difficulty of GTB versus Escape Room and Cleanup}.
One may wonder why the performance gap between MetaGrad and dual-RL is much higher in \textit{Escape Room} and \textit{Cleanup} than in GTB.
This is because the incentivization problem in GTB is conceptually easier for an incentive designer who uses conventional RL.
In GTB, agents can learn from positive rewards regardless of the tax rate, which implies that the ID always receives a learning signal that tracks the agents' behaviors.
For dual-RL, the ID receives such feedback in the \textit{next} episode, but at least it is non-zero.
However, in the other two problems, if incentives do not pass a threshold ($1+\epsilon$ to overcome the environment penalty in \textit{Escape Room}; a more complex opportunity cost for sacrificing self-interest in \textit{Cleanup}), then the agents do not receive any predefined environment rewards, which means an RL-based designer does not receive any feedback at all.
In this case, MetaGrad's knowledge of the way that agents' policy parameters change in response to the incentive function, and the use of online cross-validation to learn from the ID's returns in subsequent episodes, is crucial.

\section{Conclusion}
\label{sec:conclusion}

We proposed the use of complex simulations involving reinforcement learning agents as an \textit{in silico} experimental approach to problems of incentive design.
To tackle the issue of delayed impact of incentives, which poses difficulties for directly applying standard RL to the incentive designer, we proposed a meta-gradient approach for the incentive designer to account exactly for the agents' learning response to incentives.
The new method significantly outperforms baselines on benchmark problems and also improves the trade-off between productivity and equality in a complex simulated economy.

Beyond incentive design, one may consider the extension of ideas in this work to the context of \textit{mechanism} design, interpreted in the general sense of modifying the underlying dynamics of the environment \citep{kramar2020should} to shape agents' behavior and optimize a system-level objective.
More generally, we hope this work shows the feasibility of a path toward a data and simulation-driven approach for improving complex systems in society.







\FloatBarrier
\newpage
\bibliographystyle{ACM-Reference-Format} 
\bibliography{citation}


\onecolumn
\newpage
\setlength{\columnsep}{2pc}
\twocolumn
\appendix

\section{Additional discussion on methods}

\subsection{Meta-gradient with TRPO/PPO objectives}
\label{subsec:app_ppo}

We recapitulate the derivation of Trust Region Policy Optimization (TRPO) \citep{schulman2015trust} to show that it is compatible with meta-gradient RL.
From this, it also follows that PPO \citep{schulman2017proximal} is applicable to the upper level incentive designer problem by substituting the gradient with respect to incentive function parameters in the place of the gradient with respect to policy parameters.


TRPO considers the performance of a policy $\pihat$, parameterized by $\thetahat$, versus another policy $\pi$ parameterized by $\theta$.
Here, the ``hat'' notation $\hat{(\cdot)}$ has no relation to the ``updated policy'' in meta-gradient RL.
The TRPO objective is
\begin{align}\label{eq:trpo-objective}
    J(\pi) := E_{\pi}\left[ \sum_t \gamma^t R(s_t,a_t) \right]
\end{align}
It was shown that \citep[Equation~2]{schulman2015trust}
\begin{align}
    J(\pihat) &= J(\pi) + \Ebb_{\pihat} \left[ \sum_t \gamma^t A_{\pi}(s_t,a_t) \right] \\
    &= J(\pi) + \sum_s \rho_{\pihat}(s) \sum_a \pihat(a|s) A_{\pi}(s,a) \, ,
\end{align}
where $\rho_{\pi}$ is the discounted state visitation frequencies and $A_{\pi}$ is the advantage function under policy $\pi$.
TRPO makes a local approximation, whereby $\rho_{\pihat}$ is replaced by $\rho_{\pi}$.
One can define
\begin{align}
    L_{\pi}(\pihat) := J(\pi) + \sum_s \rho_{\pi}(s) \sum_a \pihat(a|s) A_{\pi}(s,a) \, ,
\end{align}
and derive the lower bound $J(\pihat) \geq L_{\pi}(\pihat) - C D^{\text{max}}_{\text{KL}}(\pi, \pihat)$, where $D^{\text{max}}_{\text{KL}}$ is the KL divergence maximized over states and $C$ depends on $\pi$.
The KL divergence penalty can be replaced by a constraint, so the problem becomes
\begin{align}
    &\max_{\thetahat} \sum_s \rho_{\theta}(s) \sum_a \pihat_{\thetahat}(a|s) A_{\theta}(s,a) \\
    &\text{s.t. } \bar{D}^{\theta}_{\text{KL}}(\theta, \thetahat) \leq \delta \, ,
\end{align}
where $\bar{D}^{\theta}_{\text{KL}}$ is the KL divergence averaged over states $s \sim \rho_{\theta}$.
Using importance sampling, the summation over actions $\sum_a (\cdot)$ is replaced by $\Ebb_{a \sim q} \left[ \frac{1}{q(a|s)} (\cdot) \right]$.
It is convenient to choose $q = \pi_{\theta}$, which results in:
\begin{align}
    &\max_{\thetahat} \Ebb_{s \sim \rho_{\theta}, a\sim \pi_{\theta}} \left[ \frac{\pihat_{\thetahat}(a|s)}{\pi_{\theta} (a|s)} A_{\theta}(s,a) \right] \\
    &\text{s.t. } \Ebb_{s \sim \rho_{\theta}} \left[ D_{\text{KL}}(\pi_{\theta}(\cdot |s), \pihat_{\thetahat}(\cdot |s) ) \right] \leq \delta \, .
\end{align}

During online learning, the $\thetahat$ that is optimized and the old $\theta$ are the same at each iteration, so the gradient estimate is
\begin{align}
    \Ebb_{\pi_{\theta}}\left[ \frac{\nabla_{\theta} \pi_{\theta}(a|s)}{\pi_{\theta} (a|s)} A_{\theta}(s,a) \right] \, .
\end{align}
Now, making the connection to meta-gradient RL for the incentive design problem, we note the formal equivalency between the TRPO objective $J(\pi)$ \eqref{eq:trpo-objective} and the ID's objective $J^{\text{ID}}(\eta;\hat{\thetabf})$ \eqref{eq:bilevel_objective}, and between the TRPO agent policy $\pi_{\theta}$ and the multi-agent joint policy $\pibf_{\hat{\thetabf}(\eta)}$. \footnote{Here, when used with argument $\eta$, the ``hat'' notation is interpreted in the metagradient context, where it denotes the updated policy after learning from incentives.}
Hence, we arrive at the meta-gradient based on TRPO
\begin{align}
    \Ebb_{\pibf_{\hat{\thetabf}}}\left[ \frac{\nabla_{\eta} \pibf_{\hat{\thetabf}(\eta)}(\abf|s)}{\pibf_{\hat{\thetabf}} (\abf|s)} A_{\hat{\thetabf}}(s,\abf) \right] \, ,
\end{align}
where only the updated policy $\thetahat$ appears explicitly.
This also justifies the use of the PPO loss function as the outer loss for meta-gradient RL.

\subsection{Q-learning agents}
If agents conduct Q-learning with a set of individual Q-functions $\lbrace Q_{\theta^i}(o^i,a^u) \rbrace_{i=1}^n$, we may assume that the induced policy of each agent $i$ is:
\begin{align}\label{eq:q-policy}
    \pi_{\theta^i}(a^i|o^i) := \frac{\exp(\tau Q_{\theta^i}(o^i,a^i))}{\sum_{a \in \Acal} \exp(\tau Q_{\theta^i}(o^i,a))}
\end{align}
where $\tau$ is some constant.
Agents conduct their standard Q-learning updates as
\begin{align}\label{eq:q-learning}
    \hat{\theta}^i &\leftarrow \theta^i + \alpha f(\theta^i,\eta,\taubf) \\
    f(\theta^i,\eta,\taubf) &:= \Ebb_{\taubf \sim \pibf} \left[ \nabla_{\theta^i}\left( R^{i,\text{tot}}(s,\abf,\eta) \right. \right. \\
    &+ \left. \left. \gamma \max_{a} Q'(o^i_{t+1},a) - Q_{\theta^i}(o^i_t,a^i_t) \right)^2 \right] \, .
\end{align}
We can then use \eqref{eq:ppo_gradient} with \eqref{eq:q-policy} as the agents' policies, and $\nabla_{\eta} \hat{\theta}^i$ can be computed by differentiating through \eqref{eq:q-learning}.

\subsection{Relation to hypergradients and implicit differentiation}

Hyperparameter optimization seeks the best hyperparameters $\lambda^*$ such that the validation loss $\Lcal_V$ is minimized by the model whose weights $w^*$ are obtained by minimizing the training loss $\Lcal_T$ with  $\lambda^*$.
It can be formulated as a bi-level optimization problem \citep{lorraine2020optimizing}:
\begin{subequations}
\begin{align}
    &\lambda^* = \argmin_{\lambda} \Lcal_V(\lambda, w^*(\lambda)) \\ 
    &\text{s.t. } w^*(\lambda) = \argmin_w \Lcal_T(\lambda, w)
\end{align}
\end{subequations}
This formulation requires knowing the best-response $w^*(\lambda)$ and the Jacobian $\frac{\partial w^*(\lambda)}{\partial \lambda}$.
However, it may not be necessary, or even the best method for fastest training, to know the best-response.
A $\lambda$ corresponds to a certain optimization landscape, and the best response $w^*(\lambda)$ is the minimizer of the training loss of that landscape, with an associated validation loss.
It is not clear that the best $w^*(\lambda_k)$ for each intermediate $\lambda_k$ at training iteration $k$ produces the best direction for the update $\lambda_{k+1} \leftarrow \lambda_k + \Delta \lambda$.
We only want the final set of optimal weights, and there is no obvious reason to care about the trajectory of best response weights $w^*(\lambda_1),\dotsc,w^*(\lambda_k),\dotsc$, especially when this trajectory may not lead to convergence to the minimizer of the validation loss.
In our algorithm, we unroll and differentiate through the 1-step inner optimization, without requiring convergence of the inner optimization to a best response.

\section{Experimental setup}
\label{sec:app_experiment}

\subsection{Environment details}
\label{app:environment}

\subsubsection{Escape Room}
\label{app:escape-room}

Each agent observes a 1-hot encoding of its own position in the three possible states (lever, start, and door), as well as 1-hot encodings of the positions of all other agents.
The incentive designer observes a concatenation of 1-hot encodings of all agents' positions.
Each agent has has three actions that move it to the three available states: lever, start, and door.
At each time step, the designer uses all agents' chosen actions along with the designer's global observation to compute the incentives.
Agents receive the sum of predefined environment rewards and incentives.
An agent's individual reward is zero for staying at the current state, -1 for movement away from its current state if fewer than $M$ agents move to (or are currently at) the lever, and +10 for moving to (or staying at) the door if at least $M$ agents pull the lever.
Each episode terminates when any agent successfully exits the door, or when 5 time steps elapse.


\subsubsection{Cleanup}
\label{app:cleanup}

We used the open-source implementation based on \citep{vinitsky2020,yang2020learning}, which contains the following modifications that increase the difficulty of the social dilemma compared to the original version \citep{hughes2018inequity}: rotation actions and the tagging beam are disabled, and all agents have a fixed ``upward'' orientation.
Hence, an agent must move to the river to clear waste successfully---it cannot simultaneously stay in the region where apples spawn and fire its cleaning beam toward the river---which incurs the risk that other agents exploit the opportunity to collect apples.
Each agent receives an egocentric normalized RGB image observation that spans a sufficiently large area such that the entire map is observable by that agent regardless of its position.
The incentive designer's input is a global observation of the entire RGB map and a multi-hot vector that indicates which agent(s) used their cleaning beam.

\subsubsection{Gather-Trade-Build}
\label{app:foundation}


Each agent's observation is a pair of tensor with shape 11x11x9 and a vector of size 136, consisting of a limited egocentric spatial window, its own inventories and skills, bids and asks in the market, the tax rates of the current period as well as the marginal rate corresponding to its current income so far.
This is in accord with the reverse Stackelberg game formulation, whereby agents observe the designer's chosen function.
The ID's observation is a pair of tensor with shape 14x14x8 and a vector of size 112, consisting of complete spatial world state, agents' inventories, all bids and asks, and all derived tax quantities, but does not contain agents' private skill and utility functions.

Each agent's instantaneous reward at time step $t$ is defined as
\begin{align}\label{eq:foundation_reward}
    r^i_t := u^i(x^i_t, l^i_t) - u^i(x^i_{t-1},l^i_{t-1}) \, ,
\end{align}
where $x^i_t$ is the total resources (stone and wood) and coin owned by agent $i$ at time $t$, and $l^i_t$ is the cumulative labor due to actions by agent $i$ up to time $t$.
The utility function is defined as:
\begin{align}\label{eq:utility}
    u^i(x^i_t, l^i_t) := \text{crra}(x^{i,c}_t) - l^i_t \quad \quad \text{crra}(z) := \frac{z^{1-\eta} - 1}{1 - \eta}, \quad \eta > 0
\end{align}
where $x^{i,c}_t$ is the amount of coin owned by agent $i$.

\citet{zheng2020ai} used a factored discrete action space whereby the tax planner's action is $\tau \in [0,1]^B$ where $B = 7$ is the number of tax brackets.
They designed the discrete action subspace for each bracket to be $\lbrace 0, 0.05, \dotsc, 1.0 \rbrace$.
For our meta-gradient approach, we let $r_{\eta}$ have 7 real-valued output nodes bounded in $[0,1]$ and interpret them as the tax rate for each bracket.

We applied an annealing schedule that caps the maximum marginal tax rate chosen by dual-RL, as done in previous work \citep{zheng2020ai}.
The cap increases linearly from $0.1$ to $1.0$ (i.e., no cap) within the first $8k$ episodes.


\subsection{Implementation}
\label{subsec:app_implementation}


In \textit{Escape Room}, the agent's policy network has two fully-connected (FC) layers of size $h_1$ and $h_2$ each, followed by a \texttt{softmax} output layer with size equal to the action space size.
The MetaGrad incentive designer concatenates its observation with the actions taken by all agents, passes them through two FC layers of sizes $d_1$ and $d_2$, then to a \texttt{sigmoid} output layer with size equal to the number of possible agent actions.
The dual-RL (c) incentive designer has the same architecture as in MetaGrad, except that the output layer is linear and is interpreted as the mean of a Gaussian distribution.
The dual-RL (d) incentive designer has the same architecture as the agents.

In \textit{Cleanup}, the agents have the following neural network layers, all with \texttt{ReLU} activtion.
The input image is passed through one convolutional layer with $6$ filters of kernel size $[3,3]$ and stride $[1,1]$.
This is flattened and passed through two FC layers with sizes $h_1$ and $h_2$.
For the agents' policy network, there is a final \texttt{softmax} output layer that maps to action probabilities.
The agents' value function has the same architecture as the policy, except for a linear output layer.
The incentive designer has the same convolutional layer as agents' policies. The image part of the designer's observation is passed through that layer, then through an FC layer with size $h_1$, then concatenated with the vector part of the designer's observation (the multi-hot vector indicating usage of cleaning beam), then passed through the second FC layer with size $h_2$, then finally to a \texttt{sigmoid} output layer with size $3$ (one per action type in the set $\lbrace \text{clean, collect apple, else} \rbrace$.
The dual-RL (c) designer has the same architecture as the MetaGrad designer, except for a linear output layer than is interpreted as the mean of a diagonal Gaussian distribution for its continuous action space.

In \textit{Gather-Trade-Build} without curriculum, the agents' policy network consists of the following neural network layers (all with $\texttt{ReLU}$ activation) that process the agent's observation (which has a 3D image part and a 1D vector part).
The image part of the agent's observation is passed through two convolutional layers with $6$ filters of kernel size $[5,5]$ each and stride $[1,1]$; this is flattened and passed through a fully-connected (FC) layer whose output size equals the size of the 1D vector observation; this is concatenated with the 1D vector observation, then passed through two FC layers of with 128 nodes each, then finally to a \texttt{softmax} output layer size output size equal to the discrete action space size.
The incentive designer's policy in dual-RL is exactly the same as the agents' policy architecture, except that the two FC layers have 256 nodes each.
The incentive designer in MetaGrad is the same as the dual-RL designer policy, except that the output layer has \texttt{sigmoid} activation (rather than \texttt{softmax}) and has size $7$ (for the $7$ tax brackets).
The same architecture is used for the agents' value function (which does not share parameters with the policy network), except that we used an LSTM layer of size 128 after the two FC layers.

In \textit{Gather-Trade-Build} with curriculum, the agent and designer policy and value networks are the same as the case without curriculum, except for these differences: there is an LSTM layer of size 128 after the two FC layers for the incentive designer's policy (i.e., incentive function in the case of MetaGrad, regular policy in the case of dual-RL) and value networks, and for the agents' value network.
We omitted the LSTM only for the agents' policy network to avoid complications with second-order gradients in MetaGrad.

The incentive designer in MetaGrad differentiates its objective with respect to the tax function parameters, through the agents' learning, by replicating the chain of calculations from tax $T(z)$ \eqref{eq:tax_formula} to the agents' final instantaneous reward $r^i_t$ \eqref{eq:foundation_reward} within the overall computational graph.
All quantities required to compute tax and total reward are saved in the designer's episode buffer to use in the meta-gradient step.

\subsection{Hyperparameters}
\label{subsec:app_hyperparameters}

We used random uniform sampling with successive elimination for hyperparameter search for all methods.
We start with a batch of $n_{\text{batch}}$ tuples, where each tuple is a combination of hyperparameter values with each value sampled either log-uniformly from a continuous range or uniformly from a discrete set.
We train independently with each tuple for $n_{\text{episode}}$ episodes, eliminate the lower half of the batch based on their final performance, then initialize the next set of $n_{\text{episode}}$ episodes with the current models for the remaining tuples.
We use the hyperparameters of the last surviving model.
For \textit{Escape Room}, we used $n_{\text{batch}} = 128$ and $n_{\text{epsiodes}} = 8000$.
For \textit{Cleanup}, we used $n_{\text{batch}} = 128$ and $n_{\text{epsiodes}} = 2800$.
For \textit{GTB}, we used $n_{\text{batch}} = 128$ and $n_{\text{epsiodes}} = 500$.

Let $c_{\text{entropy}}$ denote the policy entropy coefficient, $\alpha_{\theta}$ the agent's policy learning rate, $\alpha_{v,\text{agent}}$ the agent's value function learning rate, $\alpha_{\text{ID}}$ the incentive designer's learning rate (dual-RL's policy, MetaGrad's incentive function), $\alpha_{v,\text{ID}}$ the incentive designer's value function learning rate, $c_v$ the value function target update rate (i.e., $\theta'_v \leftarrow c_v \theta_v + (1-c_v) \theta'_v$, where $\theta'_v$ are parameters of a separate target network and $\theta_v$ are parameters of the main value network).
For MetaGrad in \textit{Escape Room}, we used a separate optimizer for the cost part of the ID's objective, with learning rate denoted by $\alpha_{\text{cost}}$.

The hyperparameter ranges used for tuning are the following.
\begin{itemize}[leftmargin=7pt]
    \item For \textit{Escape Room}: $c_{\text{entropy}} \in (0.01, 10.0)$, 
$\alpha_{\theta} \in (10^{-5}, 10^{-3})$, 
designer $c_{\text{entropy}} \in (0.01, 10.0)$, 
$\alpha_{\text{ID}} \in (10^{-5}, 10^{-2})$, 
agent first hidden layer $h_1 \in \lbrace 64, 128 \rbrace$, 
agent second hidden layer $h_2 \in \lbrace 16, 32, 64 \rbrace$, 
designer first hidden layer $d_1 \in \lbrace 64, 128 \rbrace$, 
designer second hidden layer $d_2 \in \lbrace 16, 32, 64 \rbrace$, 
$\alpha_{\text{cost}} \in (10^{-5}, 10^{-3})$.
\item For \textit{Cleanup}: agent $c_{\text{entropy}} \in (10^{-3}, 1.0)$, 
agent $\epsilon_{\text{start}} \in \lbrace 0.5, 1.0 \rbrace$, 
agent $\epsilon_{\text{end}}\in \lbrace 0.05, 0.1 \rbrace$, 
agent $\epsilon_{\text{div}} \in \lbrace 100, 1000, 5000 \rbrace$, 
$\alpha_{\theta} \in (10^{-5}, 10^{-3})$
$\alpha_{v, \text{agent}} \in (10^{-5}, 10^{-3})$,
agent $c_v \in (10^{-3}, 1)$,
first FC layer $h_1 \in \lbrace 64, 128, 256 \rbrace$, 
second FC layer $h_2 \in \lbrace 64, 128, 256 \rbrace$,
designer $c_{\text{entropy}} \in (10^{-3}, 1.0)$,
$\alpha_{\text{ID}} \in (10^{-5}, 10^{-3})$,
$\alpha_{v,\text{ID}} \in (10^{-5}, 10^{-3})$,
designer PPO $\epsilon \in (0.01, 0.5)$,
designer $c_v \in (10^{-3}, 1)$
\item For \textit{Gather-Trade-Build}: agent $c_{\text{entropy}} \in (10^{-3}, 10.0)$, 
$\alpha_{\theta} \in (10^{-5}, 10^{-3})$, 
$\alpha_{v, \text{agent}} \in (10^{-5}, 10^{-3})$, 
agent PPO $\epsilon \in (0.01, 0.5)$, 
agent $c_v \in (10^{-3}, 1)$, 
designer $c_{\text{entropy}} \in (10^{-3},10.0)$, 
$\alpha_{\text{ID}} \in (10^{-5}, 10^{-3})$, 
$\alpha_{v,\text{ID}} \in (10^{-5}, 10^{-3})$, 
designer PPO $\epsilon \in (0.01, 0.5)$, 
designer $c_v \in (10^{-3}, 1.0)$.
\end{itemize}

In \textit{Escape Room}, we used discount $\gamma = 0.99$, gradient descent for agents, and AdamOptimizer \citep{abadi2016tensorflow} for the ID.
In \textit{Cleanup}, we used discount $\gamma = 0.99$, dual-RL designer GAE $\lambda = 0.99$, gradient descent for agents, and AdamOptimizer for the ID.
In \textit{Gather-Trade-Build}, the fixed hyperparameters (i.e., not part of tuning) are: GAE $\lambda = 0.98$ for both agents and the ID, discount factor $\gamma = 0.99$, agent gradient clipping by $10.0$ (for dual-RL and US federal only), truncation of an episode rollout to subsequences of length $50$ for LSTMs.
We used AdamOptimizer for both agents and the ID.
For the case without curriculum, and for MetaGrad with curriculum, we applied an exploration lower bound on the agents' policies such that the actual policy is $\pi(a|s) = (1-\epsilon) \hat{\pi}(a|s) + \epsilon / \lvert \Acal \rvert$, where $\lvert \Acal \rvert$ is the size of the discrete action space and $\epsilon$ linearly decreases from $\epsilon_{\text{start}}$ to $\epsilon_{\text{end}}$ over $\epsilon_{\text{div}}$ episodes.
We did not apply this exploration lower bound to dual-RL and US federal in the curriculum case as it was not used in previous work \citep{zheng2020ai}.

\begin{table*}[t]
    \centering
    \caption{Hyperparameters in Escape Room.}
    \label{tab:hyperparam-er}
    \begin{tabular}{lrrrrrr}
        \toprule
        \multicolumn{1}{c}{} &
        \multicolumn{3}{c}{ER$(5,2)$} & \multicolumn{3}{c}{ER$(10,5)$}\\
        \cmidrule(r){2-4}
        \cmidrule(r){5-7}
        Parameter & MetaGrad & dual-RL (c) & dual-RL (d) & MetaGrad & dual-RL (c) & dual-RL (d) \\
        \midrule
        Agent $c_{\text{entropy}}$ & $0.0166$ & $7.07$ & $0.386$ & $0.0166$ & $0.345$ & $0.0166$ \\
        $\alpha_{\theta}$ & $9.56\cdot10^{-5}$ & $7.70\cdot10^{-4}$ & $3.41\cdot10^{-4}$ & $9.56\cdot10^{-5}$ & $9.05\cdot10^{-4}$ & $9.56\cdot10^{-5}$\\
        Designer $c_{\text{entropy}}$ & - & - & $0.488$ & - & - & $0.148$ \\
        $\alpha_{\text{cost}}$ & $6.03\cdot 10^{-5}$ & - & - & $6.03\cdot 10^{-5}$ & - & - \\
        $\alpha_{\text{ID}}$ & $7.93\cdot 10^{-4}$ & $1.17\cdot10^{-4}$ & $2.47\cdot10^{-3}$ & $7.93\cdot 10^{-4}$ & $1.63\cdot10^{-4}$ & $7.07\cdot10^{-3}$ \\
        $h_1$ & $64$ & $64$ & $128$ & $64$ & $64$ & $64$ \\
        $h_2$ & $64$ & $32$ & $64$ & $64$ & $32$ & $64$\\
        $d_1$ & $64$ & $64$ & $128$ & $64$ & $128$ & $64$ \\
        $d_2$ & $32$ & $64$ & $64$ & $32$ & $64$ & $32$ \\
        \bottomrule
    \end{tabular}
\end{table*}

\begin{table*}[h]
    \centering
    \caption{Hyperparameters in Cleanup.}
    \label{tab:hyperparam-cleanup}
    \begin{tabular}{lrr}
        \toprule
        Parameter & MetaGrad & dual-RL (c) \\
        \midrule
        Agent $c_{\text{entropy}}$ & $0.129$ & $0.173$ \\
        Agent $\epsilon_{\text{start}}$ & $1.0$ & $1.0$ \\
        Agent $\epsilon_{\text{end}}$ & $0.05$ & $0.1$ \\
        Agent $\epsilon_{\text{div}}$ & $100$ & $100$ \\
        $\alpha_{\theta}$ & $6.73\cdot10^{-4}$ & $6.23\cdot10^{-4}$ \\
        $\alpha_{v, \text{agent}}$ & $5.54\cdot10^{-4}$ & $4.46\cdot10^{-4}$ \\
        Agent $c_{v}$ & $7.93\cdot10^{-3}$ & $0.292$ \\
        Designer $c_{\text{entropy}}$ & - & $0.23$ \\
        $\alpha_{\text{ID}}$ & $1.24\cdot10^{-5}$ & $1.15\cdot10^{-5}$ \\
        $\alpha_{v, \text{ID}}$ & $2.4\cdot10^{-5}$ & $1.72\cdot10^{-5}$ \\
        Designer PPO $\epsilon$ & $0.0172$ & $0.0164$ \\
        Designer $c_v$ & $0.114$ & $0.846$ \\
        $h_1$ & $64$ & $64$ \\
        $h_2$ & $64$ & $64$ \\
        \bottomrule
    \end{tabular}
\end{table*}

\begin{table*}[t]
    \centering
    \caption{Hyperparameters in Gather-Trade-Build.}
    \label{tab:hyperparam-gtb}
    \begin{tabular}{lrrrrrr}
        \toprule
        \multicolumn{1}{r}{} &
        \multicolumn{3}{c}{No curriculum} & \multicolumn{3}{c}{Curriculum}\\
        \cmidrule(r){2-4}
        \cmidrule(r){5-7}
        Parameter & MetaGrad & dual-RL & US Federal & MetaGrad & dual-RL & US Federal \\
        \midrule
        Agent $c_{\text{entropy}}$ & $0.184$ & $0.282$ & $0.253$ & $0.0316$ & $0.0166$ & $0.0811$ \\
        Agent $\epsilon_{\text{start}}$ & $0.5$ & $0.5$ & $0.5$ & $0.5$ & - & - \\
        Agent $\epsilon_{\text{end}}$ & $0.05$ & $0.05$ & $0.05$ & $0.05$ & - & - \\
        Agent $\epsilon_{\text{div}}$ & $5k$ & $5k$ & $5k$ & $5k$ & - & - \\
        $\alpha_{\theta}$ & $3.30\cdot10^{-3}$ & $3.92\cdot10^{-4}$ & $1.39\cdot10^{-4}$ & $1.09\cdot10^{-5}$ & $3.33\cdot10^{-4}$ & $1.32\cdot10^{-4}$ \\
        $\alpha_{v, \text{agent}}$ & $1.67\cdot10^{-5}$ & $3.74\cdot10^{-4}$ & $5.0\cdot10^{-5}$ & $1.14\cdot10^{-5}$ & $5.58\cdot10^{-5}$ & $8.87\cdot10^{-5}$ \\
        Agent PPO $\epsilon$ & $0.01$ & $0.0201$ & $0.166$ & $0.0308$ & $0.084$ & $0.034$ \\
        Agent $c_{v}$ & $0.0126$ & $2.80\cdot10^{-3}$ & $4.36\cdot10^{-3}$ & $0.021$ & $0.221$ & $ 0.702$ \\
        Designer $c_{\text{entropy}}$ & - & $0.203$ & - & - & $0.33$ & - \\ 
        $\alpha_{\text{ID}} $ & $1.82\cdot10^{-5}$ & $7.80\cdot10^{-4}$ & - & $2.03\cdot10^{-4}$ & $4.9\cdot10^{-5}$ & - \\
        $\alpha_{v, \text{ID}}$ & $8.64\cdot10^{-4}$ & $2.30\cdot10^{-4}$ & - & $10^{-5}$ & $5.14\cdot10^{-4}$ & - \\
        Designer PPO $\epsilon$ & $0.382$ & $0.0216$ & - & $0.0394$ & $0.0295$ & - \\
        Designer $c_v$ & $0.0158$ & $1.90\cdot10^{-3}$ & - & $0.0573$ & $0.0178$ & - \\
        Designer $\epsilon_{\text{start}}$ & - & $0.5$ & - & - & - & - \\
        Designer $\epsilon_{\text{end}}$ & - & $0.05$ & - & - & - & - \\
        Designer $\epsilon_{\text{div}}$ & - & $5k$ & - & - & - & - \\        
        \bottomrule
    \end{tabular}
\end{table*}

\subsection{Computation resources}

Experiments were run on the following hardware: Intel(R) Core(TM) i7-4790 CPU with NVIDIA GeForce GTX 750 Ti GPU; Intel(R) Xeon(R) CPU E5-2630 v4 with NVIDIA GeForce GTX 1080 GPU.

\section{Additional results}
\label{app:additional-results}

\begin{figure}
    \centering
    \begin{minipage}{0.49\linewidth}
        \centering
        \includegraphics[width=1.0\linewidth]{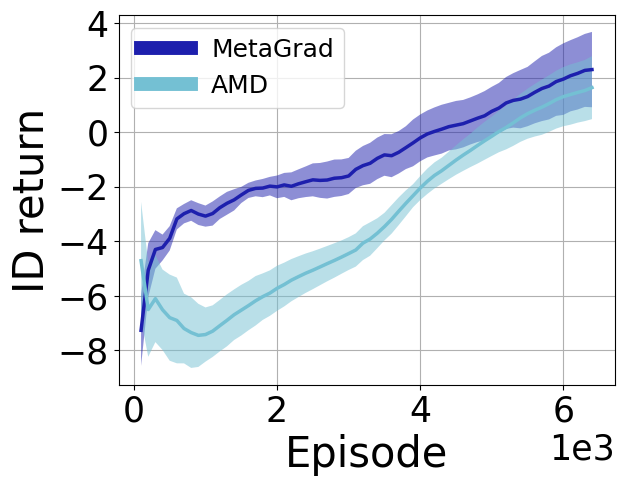}
    \caption{ER$(2,1)$}
    \label{fig:n2m1_amd}
    \end{minipage}
    \hfill
    \begin{minipage}{0.49\linewidth}
        \centering
        \includegraphics[width=1.0\linewidth]{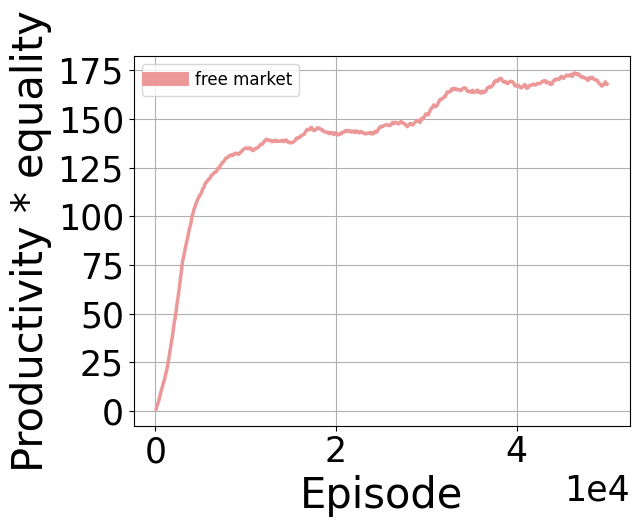}
        \caption{GTB with curriculum: Phase 1 train curve}
        \label{fig:15x15_phase1}
    \end{minipage}
\end{figure}

The central planner in \citet{baumann2020adaptive} is the centralized analogue of a LOLA agent \citep{foerster2018learning}, who learns to shape opponent behavior by anticipating their policy update.
\citet{letcher2018stable} showed analytically in certain settings that this can result in aggressive behaviors that try to force opponent compliance, especially in settings with multiple LOLA agents.
Moreover, the anticipatory behavior of a LOLA-type central planner does not benefit from the knowledge of the actual impact of incentives on recipients' behavior, as previously shown experimentally in the context of decentralized incentivization \citet{yang2020learning}.
To match the open source implementation, we used a single-layer linear policy network and actor-critic agents.
Although \citet{baumann2020adaptive} report good results in matrix games, we see in \Cref{fig:n2m1_amd} that this method learns slower than MetaGrad.
Runs were truncated after 6k episodes as the method was unstable on the temporally-extended Escape Room game.

\end{document}